\documentclass[a4paper,12pt]{article}
\usepackage{authblk}
\usepackage[square,sort,comma,numbers]{natbib}
\usepackage{soul}
\usepackage{color}
\usepackage{pifont}
\usepackage{mathptmx}
\usepackage{graphicx}
\usepackage{amsmath, amsfonts, amssymb, mathrsfs}
\usepackage{ccaption}
\usepackage{multirow}
\usepackage{fixmath}
\usepackage{tabu}
\usepackage{longtable}
\newcommand{\cmark}{\ding{51}}%
\newcommand{\xmark}{\ding{55}}
\DeclareMathOperator*{\argmax}{arg\,max}
\DeclareMathOperator*{\argmin}{arg\,min}

\title{The Optimal Dynamic Treatment Rule SuperLearner: Considerations, Performance, and Application}
\author[1]{Lina Montoya \thanks{Email address for correspondence: lmontoya@unc.edu}}
\affil[1]{University of North Carolina, Chapel Hill, Department of Biostatistics}
\author[2]{Mark van der Laan}
\affil[2]{University of California, Berkeley, Division of Biostatistics}
\author[3]{Alexander Luedtke}
\affil[3]{University of Washington, Department of Statistics}
\author[4]{Jennifer Skeem}
\affil[4]{University of California, Berkeley, Departments of Social Welfare and Public Policy}
\author[2]{Jeremy Coyle}
\author[2]{Maya Petersen}
\date{October 2021}

\begin{document}

\maketitle
\pagebreak

\begin{abstract}

The optimal dynamic treatment rule (ODTR) framework offers an approach for understanding which kinds of patients respond best to specific treatments -- in other words, treatment effect heterogeneity. Recently, there has been a proliferation of methods for estimating the ODTR. One such method is an extension of the SuperLearner algorithm – an ensemble method to optimally combine candidate algorithms extensively used in prediction problems – to ODTRs. Following the ``causal roadmap," we causally and statistically define the ODTR and provide an introduction to estimating it using the ODTR SuperLearner. Additionally, we highlight practical choices when implementing the algorithm, including choice of candidate algorithms, metalearners to combine the candidates, and risk functions to select the best combination of algorithms. Using simulations, we illustrate how estimating the ODTR using this SuperLearner approach can uncover treatment effect heterogeneity more effectively than traditional approaches based on fitting a parametric regression of the outcome on the treatment, covariates and treatment-covariate interactions. We investigate the implications of choices in implementing an ODTR SuperLearner at various sample sizes. Our results show the advantages of: (1) including a combination of both flexible machine learning algorithms and simple parametric estimators in the library of candidate algorithms; (2) using an ensemble metalearner to combine candidates rather than selecting only the best-performing candidate; (3) using the mean outcome under the rule as a risk function. Finally, we apply the ODTR SuperLearner to the ``Interventions" study, an ongoing randomized controlled trial, to identify which justice-involved adults with mental illness benefit most from cognitive behavioral therapy (CBT) to reduce criminal re-offending. 
\end{abstract}

\pagebreak

\section{Introduction}

The primary objective of a clinical trial is often to evaluate the overall, average effect of a treatment on an outcome in a given population \citep{dahabreh2016using, kent2018personalized, kosorok2019precision}. To accomplish this objective in the point treatment setting, baseline covariate, treatment, and outcome data are often collected and the average treatment effect (ATE) is estimated, quantifying the average impact of the treatment in a population. Researchers may then interpret the impact of the treatment as beneficial, neutral, or harmful. In this interpretation, the treatment’s impact is one-size-fits-all; in other words, the effect of the treatment is interpreted as the same for everyone in the study population. But, it may be the case that an intervention tends to yield better outcomes for certain kinds of people but not for others. For example, because justice-involved people with mental illness are a heterogeneous group with diverse symptoms, risk factors, and other treatment-relevant characteristics \citep{skeem2011correctional, skeem2014offenders}, assigning Cognitive Behavioral Therapy (CBT) may decrease the probability of recidivism for individuals with high risk of recidivism but not low risk of recidivism \citep{lipsey2007effects}. The ATE analysis may lead one to conclude that there is no treatment effect in a given population, when there is, in fact, a differential treatment effect for levels of variables.

Precision health aims to shift the question from “which treatment works best” to “which treatment works best \emph{for whom}?” (sometimes, it further asks: at what time? And/or at what dose? \citep{kosorok2019precision}). The point of moving towards this question is to move towards better participant outcomes. While a range of novel study designs can help to address these questions by generating data in which individualized treatment effects are unconfounded \citep{kosorok2019precision, lei2012smart, hu2006theory}, data from classic randomized controlled trials also provide a rich data source for discovering treatment effect heterogeneity. Under the assumption of no unmeasured confounding, the same methods can be applied to observational data.

One way of learning which treatment works best for whom is to estimate effects within subgroups. Following our above example within the field of criminal justice, one could split the sample into older defendants versus younger defendants, and look at the average effect of CBT on recidivism within these two age categories. Such a classic subgroup analysis helps to move a step closer to understanding the treatment that works best for whom. However, the need to restrict the number of tests performed and to pre-specify analyses limits traditional subgroup analyses to comparing intervention effects in a small set of subgroups in which heterogeneous treatment effects are expected \citep{kent2018personalized, lipkovich2017tutorial}. In practice, the participant characteristics that are most important for determining the best-suited intervention may not be clear based on background knowledge. Further, effectively predicting the type of intervention that an individual will best respond to may require accounting for a wide range of participant characteristics and complex interactions between them. For instance, identifying the individuals most likely to respond to CBT versus, for example, treatment as usual (TAU) may require considering not only risk level, but also age, educational attainment, sex, substance abuse, psychological distress, and internal motivation to adhere to treatment -- as well as various interactions between these. In summary, the challenge is to take a wide range of participant characteristics and flexibly learn how to best combine them into a strategy or rule that assigns to each participant the specific intervention that works best for him or her.

Estimating the optimal dynamic treatment rule (ODTR) for a given population offers a formal approach for learning about heterogeneous treatment effects and developing such a strategy. A dynamic treatment rule can be thought of as a rule or algorithm where the input is participant characteristics and the output is an individualized treatment choice for each person \citep{bembomvdL2007practical, van2007causal, robins1986new, chakraborty2013statistical}. An ODTR (also known as an optimal treatment regime, optimal strategy, individualized treatment rule, optimal policy, etc.) is the dynamic treatment rule that yields the best overall participant outcomes \citep{murphy2003optimal, robins2004optimal}. In our criminal justice example, a dynamic treatment rule takes as input participant characteristics such as age, criminal history, and education level and outputs a treatment decision -- either CBT or TAU. The ODTR is the dynamic treatment rule under which the highest proportion of patients are not re-arrested. It is the most effective and, if one incorporates cost or constraints on resources \citep{luedtke2016optimal}, efficient way of allocating the interventions at our disposal based on measured individual characteristics. 

There have been major advances in estimating the ODTR within the fields of statistics and computer science, with important extensions to the case where treatment decisions are made at multiple points in time. Regression-based approaches, such as Q-learning, learn the ODTR by modeling the outcome regression (i.e., the expected outcome given treatment and covariates) directly \citep{murphy2003optimal, laber2014interactive, schulte2014q, moodie2012q, qian2011performance}. Robins and Murphy developed methods of estimating the ODTR by modeling blip-to-reference functions (i.e., the strata-specific effect of the observed treatment versus control) and regret functions (i.e., the strata-specific loss incurred when given the optimal treatment versus the observed treatment), respectively \citep{murphy2003optimal, robins2004optimal, moodie2007demystifying}. Direct-estimation approaches to learning the ODTR, such as outcome weighted learning (OWL), aim to search among a large class of candidate rules for the one that yields the best expected outcome \citep{zhao2012estimating, zhang2012estimating, zhao2015new}. These are examples of broad classes of ODTR estimators; within and outside of them there has been a proliferation of methods to estimate the ODTR (see \cite{kosorok2019precision, kosorok2015adaptive, zhao2014estimation} for reviews of the state of the art in estimating ODTRs and precision medicine).

Given the vast number of methods available for estimating the ODTR, the question becomes: which approach to use? In some settings, some algorithms may work better than others. SuperLearning \citep{polley2010super} (or, more specific to prediction, stacked regression \citep{breiman1996stacked}) was originally proposed as a method for data-adaptively choosing or combining prediction algorithms. The basic idea is to define a library of candidate algorithms and  choose the candidate or the combination of candidates that gives the best performance based on $V$-fold cross-validation. This requires defining: (1) the algorithms to include in the library, (2) a parametric family of weighted combinations of these algorithms, the “metalearning" step \citep{ledell2016auc}, and (3) the choice of performance metric (i.e., risk) as the criterion for selecting the optimal combination of algorithms. Given these three requirements, one can estimate the risk for each combination of algorithms using $V$-fold cross-validation, and choose the combination with the lowest cross-validated risk. The SuperLearner framework has been implemented extensively for prediction problems \citep{polley2010super, pirracchio2014improving, pirracchio2015mortality, petersen2015super}, and has been extended to the ODTR setting \citep{luedtke2016super, coyle2017computational}. In particular, Luedtke and van der Laan showed that in the randomized controlled trial (RCT) and sequential multiple assignment randomized trial (SMART) \citep{kosorok2015adaptive, almirall2014introduction, lei2012smart} settings, under the assumption that the loss function is bounded, the ODTR SuperLearner estimator will be asymptotically equivalent to the ODTR estimator chosen by the oracle selector (that is, the ODTR estimator, among the candidate ODTR estimators, that yields the lowest risk under the true data distribution \citep{polley2010super}). This implies that the ODTR SuperLearner will asymptotically do as well as or better than any single candidate estimator in the library, provided that none of the candidate algorithms are based on correctly specified parametric models. If there is a well-specified regression in the library, the ODTR SuperLearner estimator of the ODTR will achieve near parametric rates of convergence to the true rule.

These theoretical results lay important groundwork for understanding the asymptotic benefits to using the algorithm; however, less has been published on how the ODTR SuperLearner performs in finite samples, the practical implications of key choices when implementing the algorithm, and illustrations of implementing this algorithm on real RCT data. In this paper, we provide an introduction to the implementation of the ODTR SuperLearner in the point treatment setting, and use simulations to investigate the tradeoffs inherent in these user-supplied choices and how they may differ with varying sample sizes. In particular, we examine: (1) how to select the candidate algorithms for estimating the ODTR; specifically, the costs and benefits to expanding the library to include a wider set of diverse ODTR algorithms, including simple parametric regressions versus more data adaptive algorithms, and blip-based versus direct estimation algorithms; (2) implications of the choice of parametric family for creating weighted combinations for candidate ODTR learners (i.e., choice of metalearner); and, (3) implications of the choice of risk function used to judge performance and thereby select the optimal weighted combination of candidate learners. We do this first for sample sizes 1,000, and then for a reduced sample size of 300 to investigate the implications of a smaller sample size for the relative performance of each ODTR SuperLearner configuration. Finally, we apply the ODTR SuperLearner to real data generated from the Correctional Intervention for People with Mental Illness, or ``Interventions," trial, an ongoing RCT in which justice-involved adults with mental illness were either randomized to CBT or TAU. In applying the ODTR SuperLearner to this sample, we aim to identify which people benefit most from CBT versus TAU, in order to reduce recidivism.

Of note, we refer the reader to our companion paper \citep{montoyaplaceholder} in which we describe and show performance of various estimators for the value of a given or learned treatment rule. Specifically, once one has estimated the ODTR, as demonstrated in this paper, it may be of interest to \textit{evaluate} it -- that is, to ask the causal question: how would expected outcomes have differed had everyone received treatment assigned by the optimal rule? As with the current paper, we additionally apply the techniques we outline in that paper to estimation of the value of a rule applied to the ``Interventions" Study.

The organization of this article is as follows. First, we step through the causal roadmap (as described in \cite{petersen2014causal}) for defining the true ODTR for a given population. We focus on the case in which baseline covariates are measured, a single binary treatment is randomized, and an outcome is measured. We then give a brief introduction to some estimators of the ODTR, and in particular, describe the SuperLearner approach for estimating the optimal rule that builds on Luedtke and van der Laan's work \citep{luedtke2016super}. We investigate the implications of the three sets of implementation choices outlined above in finite samples using simulations (with corresponding R code illustrating implementation of all estimators considered), and the performance under such options. Lastly, we show results for the ODTR SuperLearner algorithm applied to the ``Interventions" Study. We close with concluding remarks and future directions. In the Appendix, we provide a Notation Table with terms frequently used throughout this manuscript and our companion manuscript, in addition to definitions of each of the terms, as a reference to reader.

\section{Causal Roadmap and ODTR Framework}\label{sec2}

\subsection{Data and Causal Model}

Consider point-treatment data where $W \in \mathcal{W}$ are baseline covariates, $A \in \{0,1\}$ is the treatment, and $Y \in \mathbb{R}$ is the outcome measured at the end of the study. Our data can be described by the following structural causal model (SCM), $\mathcal{M}^F$ \citep{pearl2000causality}:
\begin{align*}
    W& =f_W (U_W ),\\
    A&=f_A (W,U_A),\\
    Y&=f_Y (W,A,U_Y), 
\end{align*}

\noindent where the full data $X = (W,A,Y)$ are endogenous nodes, $U = (U_W, U_A, U_Y) \sim P_U$ are unmeasured exogenous variables, and $f = (f_W, f_A, f_Y)$ are structural equations. If it is known that data were generated from an RCT using simple randomization with equal probability to each arm, then the above structural causal model would state that $Y$ may be affected by both $W$ and $A$, but that $W$ does not affect $A$ (as in the ``Interventions" trial); this design can be represented in the above model by letting $U_A \sim Bernoulli(p = 0.5)$ and $A=U_A$. In this point treatment setting, a dynamic treatment rule is a function $d$ that takes as input some decision function $V$ of the measured baseline covariates $W$ and outputs a treatment decision: $V \rightarrow d(V) \in \{0,1\}$. For the remainder of the paper, we consider the case where $V=W$; in other words, we consider treatment rules that potentially respond to all measured baseline covariates. However, consideration of dynamic rules based on a more restrictive set of baseline covariates is also of frequent pratical interest, allowing, for example, for consideration of dynamic rules based on measurements that can be more readily attained; all methods described extend directly to this case. We denote the set of all dynamic treatment rules as $\mathcal{D}$.

The ``Interventions" data consist of baseline covariates $W$, which include intervention site, sex, ethnicity, age, Colorado Symptom Index (CSI) score (a measure of psychiatric symptoms), level of substance use, Level of Service Inventory (LSI) score (a risk score to predict future recidivism that summarizes risk factors like criminal history, educational and employment problems, and attitudes supportive of crime), number of prior adult convictions, most serious offense, Treatment Motivation Questionnaire (TMQ) score (a measure of internal motivation for undergoing treatment), and substance use level; the randomized treatment $A$, either a manualized Cognitive Behavioral Intervention for people criminal justice system (abbreviated CBT; $A=1$) or treatment as usual (TAU), primarily psychiatric or correctional services ($A=0$); and a binary outcome $Y$ of recidivism, an indicator that the person was not re-arrested within one year after study enrollment ($Y=1$ if a participant was re-arrested within one year after enrollment; $Y=0$ otherwise). In Table \ref{table1} we show the distribution of the data.

\subsection{Target Causal Parameter}

Let $d(W)$ be a deterministic function that takes as input a vector of baseline covariates, and gives as output a treatment assignment (in this case, either $0$ or $1$). For a given rule $d$, we intervene on the above SCM to derive counterfactual outcomes:
\begin{align*}
W&=f_W (U_W ),\\
A&=d(W), \\
Y_{d(W)}&=f_Y (W,d(W),U_Y).
\end{align*}
Here, $Y_{d(W)}$ is the counterfactual outcome for a person if his/her treatment $A$ were assigned using the dynamic treatment rule $d(W)$; to simplify notation we refer to this counterfactual outcome as $Y_d$. The counterfactual outcomes for a person if he/she were assigned treatment or control are denoted $Y_1$ and $Y_0$, respectively. Together, the distribution of the exogenous variables $P_U$ and structural equations $f$ imply a distribution of the counterfactual outcomes, and the SCM provides a model for the set of possible counterfactual distributions: $P_{U,X} \in \mathcal{M}^F$. 

Our target parameter of interest in this paper is the ODTR, defined as the rule that, among all candidate rules $\mathcal{D}$, yields the best expected outcomes. Using the convention that larger values of $Y$ correspond to better outcomes, an ODTR is defined as a maximizer of $\mathbb{E}_{P_{U,X} } [Y_d] $ over all candidate rules

\begin{align}
    d^* & \in \argmax_{d\in \mathcal{D}} \mathbb{E}_{P_{U,X} } [Y_d].
\label{eq1}
\end{align}

Any such ODTR can be defined in terms of the conditional average treatment effect (CATE), namely $\mathbb{E}_{P_{U,X}}[Y_1-Y_0|W]$, which is the effect of treatment for a given value of covariates $W$. Any ODTR assigns treatment 1 and 0 to all strata of covariates for which the CATE is positive and negative, respectively. If the CATE is 0 for a particular $W$ (i.e., there is no treatment effect for that strata of $W$), the ODTR as defined above may have more than one maximizing rule and therefore may be non-unique; this is why the RHS of equation \ref{eq1} above is a set \citep{luedtke2016statistical}. An ODTR can take an arbitrary value for strata at which the CATE is 0. If we assume that assigning treatment 0 is preferable to assigning treatment 1 in the absence of a treatment effect, then we would prefer the following ODTR as a function of the CATE:

\[d^*(W)\equiv \mathbb{I}\Big{[}\mathbb{E}_{P_{U,X}}[Y_1-Y_0|W] > 0\Big{]}.\]

In other words, if a person's expected counterfactual outcome is better under treatment versus no treatment given his or her covariate profile, then assign treatment; otherwise, assign control. A person’s counterfactual outcome under the ODTR is $Y_{d^*}$, and the expected outcome had everyone received the treatment assigned by the ODTR is $\mathbb{E}_{P_{U,X}}[Y_{d^*}]$.

Following our applied example, $Y_1$, $Y_0$, and $Y_d$ are the counterfactual outcomes for a person if he/she were given CBT, TAU, and either CBT or TAU based on the rule $d$, respectively; here, $d^*$ is the rule for assigning CBT versus TAU using participants' covariates that would yield the highest probability of no re-arrest, $\mathbb{E}_{P_{U,X}}[Y_{d^*}]$.

\subsection{Identification and Statistical Parameter}

We assume that our observed data were generated by sampling $n$ independent observations $O_i \equiv (W_i, A_i, Y_i)$, $i = 1, \ldots, n$, from a data generating system described by $\mathcal{M}^F$ above (e.g., the ``Interventions" study consists of 441 independent and identically distributed, or i.i.d., observations of $O$). The density of $O$ can be factorized as:

\[p_0(O)=p_{W,0}(W)g_0(A|W)p_{Y,0}(Y|A,W),\]

\noindent where $p_{W,0}$ is the true density of $W$; $g_0$ is the true conditional probability of $A$ given $W$, or the treatment mechanism; $p_{Y,0}$ is the true conditional density of $Y$ given $A$ and $W$. The distribution of the data $P_0$ is an element of the statistical model $\mathcal{M}$, which in our RCT example is semi-parametric. Further, if the data are generated from an RCT design, as in the ``Interventions" study, then the true $g_0$ is known, and the backdoor criteria (with the implied randomization assumption \citep{pearl2000causality, robins2009estimation}), $Y_d \perp A|W$ for all $d \in \mathcal{D}$, and the positivity assumption, $Pr\Big{(}\min_{a\in \{0,1\}} g_0 (A=a | W)>0\Big{)} =1$, hold by design; in an observational data setting the randomization assumption requires measurement of a sufficient set of baseline covariates, and the positivity assumption may also pose greater challenges \citep{petersen2012diagnosing}. 

Define $Q(a,w)\equiv E[Y|A=a,W=w]$ for all $a \in \{0,1\}, w \in \mathcal{W}$. Under the above assumption, $\mathbb{E}_{P_{U,X}}[Y_d]$ (a parameter of the counterfactual distribution) is identified as $\mathbb{E}_0[Q_0(A=d(W), W)]$ (a parameter of the observed distribution) for any candidate rule $d$. Thus, the ODTR is identified by

\[d_0^* \in \argmax_{d \in \mathcal{D}} \mathbb{E}_0[Q_0(A=d(W), W)].\]

In addition, the CATE is identified as $Q_0(1,W)-Q_0(0,W)$, where the latter is sometimes referred to as the blip function $B_0(W)$. Then, the true optimal rule can also be defined as a parameter of the observed data distribution using the blip function:

\[d_0^*(W) \equiv \mathbb{I}[B_0 (W)>0].\]

Analogous to the definition of the ODTR as a function of the CATE, in words, the blip function essentially says that if treatment for a type of participant $W=w$ is effective (i.e., greater than 0), treat that type of person. If not, do not treat him/her. If all participants were assigned treatment in this way, then this would result in the highest expected outcome, which is the goal.

\section{Estimation of the ODTR}

We denote estimators with a subscript $n$, so that, for example, an estimator of the true ODTR $d^*_0$ is $d^*_n$. Estimates are functions of $P_n$, which is the empirical distribution that gives each observation weight $\frac{1}{n}$. Further, if $V$-fold cross-validation is employed, the empirical data are uniformly and at random split into $V$ mutually exclusive sets. For set indices $v \in \{1,...,V\}$, each set of data serves as a validation set; the complement is its training set. Let $P_{n,v}$ be the empirical distribution of the validation sample $v$, and $P_{n,-v}$ be the empirical distribution of the complementary training set.

In what follows, we briefly describe examples of common methods for estimating the ODTR. We first describe methods that estimate the ODTR via an estimate of the blip function. We then describe methods that directly estimate a rule that maximizes the mean outcome. 

\subsection{Blip-based Approaches}

Blip-based approaches aim to learn the blip, which implies an ODTR. A benefit of doing this is that one can look at the distribution of the predicted estimates of the blip for a given sample. Having the blip distribution allows one to identify the patients in a sample who benefit most (or least, or little) from treatment. Additionally, estimating the blip function can allow for estimating the ODTR under resource constraints; for example, an ODTR in which only $k$\% of the population can receive treatment \citep{luedtke2016optimal}. Below we illustrate two methods of estimating the ODTR by way of the blip function (i.e., blip-based estimators of the ODTR).

\paragraph{Single stage Q-learning}
A plug-in estimator naturally follows from the above definition of the optimal rule. One can build the estimate $Q_n(A,W)$ of $Q_0(A,W)$ using any regression-based approach for estimating an outcome regression and predict at $Q_n(1,W)$ and $Q_n(0,W)$. This provides an estimate of the blip: $B_n(W)=Q_n(1,W)-Q_n (0,W)$, which implies an estimate of the optimal rule: $d_n^*(W)=\mathbb{I}[B_n (W)>0]$. 

\paragraph{Estimating the blip function}

Consider the double-robust pseudo-outcome:

\[D(Q,g)(O)=\frac{2A-1}{g(A|W)}[Y-Q(A,W)]+Q(1,W)-Q(0,W).\]

\noindent Importantly, $\mathbb{E}_0 [D(Q,g)(O)|W]=B_0 (W)$ if  $Q=Q_0$ or $g=g_0$. Using this result, one could estimate the blip by regressing the pseudo-outcome $D_n(Q_n,g_n)$ (which we abbreviate from here on as $D_n$) on $W$ using any regression-based approach. As in the previous method, this estimate of the blip implies an estimate of the optimal rule $d_n^*(W)=\mathbb{I}[B_n (W)>0]$.

\subsection{Direct Estimation Approaches for Maximizing the Expected Outcome}

Instead of estimating the blip function, which implies an ODTR, one could estimate the ODTR directly by selecting a rule $d$ that maximizes the estimated $\mathbb{E}_{U,X}[Y_d]$. Below we illustrate outcome weighted learning (OWL) --  one example of a direct-estimation method for the ODTR.

\paragraph{Single stage outcome weighted learning}
We briefly describe the general concept of outcome weighted learning here, but refer to \citet{zhao2012estimating} and \citet{rubin2012statistical} for a more thorough explanation. The optimal rule defined above as a function of $P_0$ could equivalently be written as an inverse probability of treatment weighted (IPTW) estimand:

\[d_0^* \in \argmax_{d \in \mathcal{D}}  \mathbb{E}_0[Q_0(A=d(W),W)]=  \argmax_{d \in \mathcal{D}} \mathbb{E}_0\Bigg{[}\frac{Y}{g_0(A|W)}\mathbb{I}[A=d(W)]\Bigg{]}.\]

Written this way, estimating $d_0^*$ could be regarded as a classification problem, where the weighted outcome $\frac{Y}{g_0(A|W)}$ helps us learn what kind of patients should get treatment: if a certain kind of patient $W=w$ has large weighted outcomes and they were treated according to candidate rule $d$, future patients with that covariate profile should be treated using that rule. Conversely, the smaller the weighted outcome among patients $W$ who were treated according to $d$, the larger the ``misclassification error" and the less likely those kinds of patients should be treated according to $d$. This maximization problem is equivalent to the following minimization problem:
\begin{align}\label{OWLest1}
d_0^* & \in \argmin_{d \in \mathcal{D}} \mathbb{E}_0\Bigg{[}\frac{Y}{g_0(A|W)}\mathbb{I}[A \neq d(W)]\Bigg{]}.
\end{align}
Now, if patients $W=w$ who did not follow the rule $d$ have large weighted outcomes (and thus larger ``misclassification error"), those kinds of patients should be given the opposite treatment that $d$ proposes. Note that in the RCT setting, if one uses the known $g_0$ and if treatments are given with equal probability, then this reduces to finding the rule that minimizes the mean outcome among patients who did not follow the rule. Equation \ref{OWLest1} could alternatively be written as a minimization problem for, instead of a rule $d$, a function $f$:
\begin{align}\label{OWLest2}
    f_0^*& \in \argmin_{f\in \mathcal{F}}  \mathbb{E}_0\Bigg{[}\frac{Y}{g_0(A|W)}I\Big{[}A \neq \frac{sign(f(W))+1}{2}\Big{]}\Bigg{]}.
\end{align}  
where $sign(x)=-1$ if $x \leq 0$ and $sign(x)=1$ if $x > 0$. Under the true data distribution $P_0$, $f_0$ is the blip function, $B_0$. In order to solve this minimization problem using data, we can use a plug-in estimator of (\ref{OWLest2}). However, since it is a 0-1 function (in particular, it is discontinuous and non-convex), one could use a convex surrogate function to approximate it, to instead minimize:
\begin{align}\label{OWLest3}
    f_n^*(W) & \in \argmin_{f\in \mathcal{F}} \frac{1}{n} \sum_{i=1}^n \frac{Y_i}{g_n (A_i|W_i)}\Phi((1-(2A_i-1)f(W_i))+ \lambda \left\lVert f \right\rVert^2,
\end{align}
where $\Phi(t)$ is a surrogate loss function (e.g., hinge loss, exponential loss, logistic loss), $\left\lVert f \right\rVert$ is a norm of $f$, and $\lambda$ is a penalization parameter on $f$ to avoid overfitting of the rule. This can also be generalized with the IPTW function replaced by the augmented IPTW \citep{luedtke2016super, rubin2012statistical}. Once $f_n^*$ is found as the solution to equation (\ref{OWLest3}), the estimated ODTR is: 

\[d_n^*(W)=sign(f_n^* (W)).\]

\subsection{SuperLearner to Estimate ODTR}

The overarching goal of SuperLearner is to let a user-supplied library of candidate algorithms, such as specific implementations of the general approaches described above, ``team up" to improve estimation of the ODTR. In order to implement the ODTR SuperLearner, there are three user-supplied decisions one must make. First, one must build the library of candidate algorithms to include. These could include algorithms for estimating the blip function (which imply a rule), algorithms that search for the ODTR directly (such as OWL estimators), static rules that determine treatment regardless of covariates, or combinations of the above classes of algorithms. Second, in what is sometimes referred to as the metalearning step, one can either implement a SuperLearner that chooses one algorithm out of the library of candidate algorithms to include (i.e., ``discrete" SuperLearner), or a SuperLearner that is a combination the candidate algorithms (i.e., ``continuous" SuperLearner). For the latter, one again has a choice of metalearner; we consider weighted convex combinations of candidate estimators of the blip and combinations of estimates of the rules themselves (through a weighted ``majority vote"). Finally, one must choose the risk function used to judge the performance of the weighted combinations of algorithms, estimated using $V$-fold cross validation (noting that $V$-fold cross validation is appropriate in our setting of i.i.d. data; other cross-validation schemes can be considered in different settings). Here, we consider two risk functions: the mean-squared error (MSE) and the mean outcome under the candidate rule. 

The steps for implementing the ODTR SuperLearner are as follows; they closely follow the implementation of the canonical SuperLearner for regression \citep{polley2010super}:
\begin{enumerate}
\item Choose $J$ candidate algorithms for estimating the optimal rule $d_{n,j}(W)$ for $j=1,...,J$. Candidates can include approaches based on estimating the blip (where the $j^{th}$ candidate estimate of the blip is denoted $B_{n,j}(W)$). Candidates can include, for example, different regression specifications of $D_n$  on $W$, where the candidate regressions might consist of a parametric linear regression (corresponding to a classic approach of fitting a parametric outcome regression on $A$ and $W$ and an interaction term) as well as more flexible machine learning type approaches such as neural networks \citep{ripley1996pattern}, multivariate adaptive regression splines \citep{friedman1991multivariate}, or recursive partitioning and regression trees \citep{breiman2017classification}. Candidate algorithms might also include approaches for estimating the optimal rule directly, such as an OWL estimator. Finally, the static treatment rules that treat all patients or treat no patients, regardless of their covariate values, can also be included as candidates. Inclusion of both simple parametric regression estimators, as well as static rules such as treating all and treating none, is important to allow for the possibility that the underlying true ODTR may in fact be simple (or well-approximated by a simple rule), and providing less aggressive candidates in the SuperLearner library can help protect against overfitting in finite samples. 
\item Split the data into $V$ exhaustive and mutually exclusive folds. Let each fold in turn serve as the validation set and the complement data as the training set.
\item Fit each of the $J$ candidate algorithms on the training set. Importantly, candidate algorithms might depend on nuisance parameters, and those nuisance parameters should be fit on the training set, as well. For example, if a candidate algorithm regresses $D_n$ on $W$ to estimate the blip (which implies an ODTR), then $Q$ and $g$ should be fit and predicted on the training set, and then plugged into $D$ to fit that candidate algorithm on the same training set (this is also called ``nested" cross-validation, described in detail by \cite{coyle2017computational}). 
\item Predict the estimated blip or the treatment assigned under the estimated ODTR for each observation in the validation set for each algorithm, based on the corresponding training set fit.
\item Choose to either implement the discrete SuperLearner, which selects one algorithm out of the candidate algorithms, or the continuous SuperLearner, which creates a weighted average of the candidate algorithms. 
\begin{enumerate}
    \item Continuous SuperLearner. Create different convex combinations of the candidate blip or treatment rule estimates that were predicted on the validation set (i.e., convex combinations of the predictions from the previous step). Formally, define an estimator of $B_{n,\alpha}(W)$ or $d_{n,\alpha}(W)$ as a convex combination of the candidate algorithms (indexed by $j$); each convex combination of algorithms is indexed by a weight vector $\alpha$. A given convex combination of blip estimates is denoted as:
    
\[B_{n,\alpha}(W)=\sum_j \alpha_j B_{n,j}(W), \alpha_j \geq 0 \forall j, \sum_j \alpha_j = 1.\] 

Alternatively, the predicted treatments under the candidate ODTRs can be combined as a weighted ``majority vote" of the convex combination of the candidate rules:

\[d_{n,\alpha}(W) = \mathbb{I}\Big{[} \sum_j \alpha_j d_{n,j}(W) > \frac{1}{2}\Big{]}, \alpha_j \geq 0 \forall j, \sum_j \alpha_j = 1.\]

Here, $\alpha_1,...,\alpha_J$ are non-negative and sum to 1.

    \item Discrete SuperLearner. The discrete SuperLearner, which chooses only one candidate algorithm, can be thought of as a special case of the continuous SuperLearner, where algorithms are still combined using a convex combination, but each algorithm weight $\alpha_j$ must be either $0$ or $1$. Such an approach may be particularly advantageous when sample size is small:
    
    \[B_{n,\alpha}(W)=\sum_j \alpha_j B_{n,j}(W), \alpha_j \in \{0,1\} \forall j, \sum_j \alpha_j = 1\] 
    
    \[d_{n,\alpha}(W) = \mathbb{I}\Big{[} \sum_j \alpha_j d_{n,j}(W) > \frac{1}{2}\Big{]}, \alpha_j \in \{0,1\}  \forall j, \sum_j \alpha_j = 1.\]
    
    Here, $\alpha_1,...,\alpha_J$ are either $0$ or $1$ and sum to $1$.
    
\end{enumerate}
\item Calculate an estimated risk within each validation set for each combination of algorithms (i.e., for each convex combination indexed by a particular value for $\alpha$). Here, we discuss two choices of risk functions for step (5) above, and their corresponding validation set-specific estimates. First, the mean-squared error risk, which we will refer to as $R_{MSE}$, is defined, for a given candidate blip function $B_{\alpha}(W)$, as:

\[R_{0,MSE}(Q,g,B)(O) = \mathbb{E}_0[(D(Q,g)(O)-B_{\alpha}(W))^2],\] 

where the risk function is indexed by nuisance parameters $Q(A,W)$ and $g(A|W)$. If either $Q=Q_0$ or $g=g_0$, then $R_{0,MSE}(Q,g,B) = \mathbb{E}_0\left[\left(B_0(W)-B_\alpha(W)\right)^2\right]$, which also equals $\mathbb{E}_{P_{U,X}}\left[\left(Y_1-Y_0-B_\alpha(W)\right)^2\right]$,  is identified as $R_{0,MSE}$, and the risk function is minimized by the true value of the blip. The validation set $v$-specific estimated risk of a candidate blip estimate, $R_{v,n,MSE}$, is calculated by first computing the estimated pseudo-outcome $D_{n,v}(O)$ in the validation set using estimates of the $Q$ and $g$ fit on the corresponding training set, i.e., $D_{n,v}(Q_{n,v},g_{n,v})(O)$. The risk for each candidate blip estimate indexed by a given $\alpha$ is then estimated by taking the empirical mean of the squared difference between this estimated pseudo-outcome and the alpha-specific blip estimate based on the training data, applied to individuals in the validation set:

\[R_{v,n,MSE}=\frac{1}{n_v}\sum_{i \in Val(v)} \left( D_{n,v}(Q_{n,v},g_{n,v})(O_i) - B_{n,\alpha,v}(W_i) \right)^2,\]

where $n_v$ is the number of persons in validation set $v$, $Val(v)$ is the indices $i$ for which $O_i$ is in the validation set, and  $B_{n,\alpha,v}$ denotes an alpha-specific blip estimate, fit on the training data training set. Of note, because this risk function is a function of the blip estimate, use of this risk function restricts the library of candidate ODTR estimators to those based on estimating the blip. 

The second risk function, which we call $R_{E[Y_d]}$, uses the expected rule-specific outcome as criterion:

\[R_{0,E[Y_d]}(d)(O)=-\mathbb{E}_{P_{U,X}}[Y_d] = -\mathbb{E}_0[Q_0(A=d(W),W)],\]

For a given estimate $d_n$, we estimate its cross validated risk with the cross-validated targeted maximum likelihood estimator (CV-TMLE) of $\frac{1}{V}\sum_{v=1}^V\mathbb{E}_{P_{U,X}}[Y_{d_{n,v}}]$ with $d_{n,v}$ being the estimator fit on the training sample $P_{n,-v}$. We describe this CV-TMLE in more detail in \citep{montoyaplaceholder}; briefly, this involves fitting $Q_0$ and $g_0$ on the training set (i.e., $Q_{n,v}$ and $g_{n,v}$), computing a TMLE update of $Q_{n,v}$ on the validation sample using $g_{n,v}$ (i.e., $Q^*_{n,v}$), evaluating the empirical mean over the validation sample of $Q^*_{n,v}(d_{n,v}(W),W)$, and averaging the latter across all splits $v = 1,...,V$. 
Intuitively, SuperLearner aims to selects the combination of treatment rule algorithms that minimizes a cross-validated empirical risk, so it makes sense to have that risk be the negative of the expected outcome, such that SuperLearner chooses the combination of algorithms that maximizes the expected outcome, since that is the ultimate goal of the ODTR. Candidate estimators for the SuperLearner that use the expected mean outcome under the rule as the risk function can include both blip estimators that imply treatment rules as well as direct estimators of the treatment rules. When the expected rule-specific outcome is chosen as the risk function, a further practical choice is how to estimate this quantity; while we focus here on a CV-TMLE due to its favorable theoretical properties (double robustness, semi-parametric efficiency, and greater protection against overfitting through the use of sample splitting \citep{van2011targeted}), one can use any estimator of treatment specific mean outcomes to estimate this quantity \citep{bembomvdL2007practical, van2011targetedint, van2007causal}.

\item Average the risks across the validation sets resulting in one estimated cross validated risk for each candidate convex combination (i.e., each possible choice of $\alpha$).
\item Choose the estimator (i.e., the convex combination $\alpha$) that yields the smallest cross-validated empirical risk. Call this “best” weighting of the algorithms $\alpha_n$.
\item Fit each candidate estimator $B_{n,j}(W)$ of the blip or $d_{n,j}(W)$ of the optimal rule on the entire data set. Generate predictions for each candidate algorithm individually, and then combine them using the weights $\alpha_n$ obtained in the previous step. This is the SuperLearner estimate of the ODTR, where $d_{n,B}^*(W)= \mathbb{I}[B_{n,\alpha_n}(W)>0]$ or $d_{n,d}^*(W)=d_{n,\alpha_n}(W)$ directly.
\end{enumerate}

We summarize the practical implications of 3 key choices for implementing ODTR SuperLearner here and in Table \ref{table0}.

\begin{itemize}
    \item The first choice is the selection of candidate algorithms. For illustration, we consider having a library with only blip function estimators (called ``Blip only" library) or a library with blip estimators, direct-estimation estimators, and static treatment rules that treat everyone or no one (called ``Full" library). If one chooses to include direct-search estimators of the ODTR or static rules (i.e., functions that do not output a blip estimate in the process), then one is constrained to using the vote-based metalearner and $R_{E[Y_d]}$ risk function, because the blip-based metalearner and $R_{MSE}$ risk function both rely on estimates of the blip for combining and choosing the algorithms, respectively. 
    \item The second choice concerns how to combine algorithms, or the metalearner. We consider either the metalearner that (a) selects only one candidate algorithm (called ``Discrete"), (b) uses a weighted average to combine predicted blip estimates and directly plugs those into the risk (called ``Blip-based"), or (c) uses a weighted average to combine predicted treatments under the candidate combinations of rules and creates a weighted majority vote of these candidate rules as input into the risk (called ``Vote-based").
    \item The third choice is that of the risk function -- either the MSE ($R_{MSE}$) or the mean outcome under the candidate rule ($R_{E[Y_d]}$). For the second and third choices, if one uses the vote-based metalearner, then the $R_{MSE}$ risk cannot be used because $R_{MSE}$ requires an estimate of the blip to choose the best algorithm, and the vote-based metalearner does not output an estimate of the blip.
\end{itemize}

\begin{center}
\begin{table}[htb]
\scalebox{0.8}{
\begin{tabular}{|c|c|c|c|c|c|c|c|c|c|c|c|c|}
\hline
\textbf{Choice 1: Library}                      & \multicolumn{6}{c|}{\textbf{Blip only}}                                                                                                                 & \multicolumn{6}{c|}{\textbf{Full}}                                                                                                                      \\ \hline
\multirow{2}{*}{\textbf{Choice 2: Metalearner}} & \multicolumn{2}{c|}{\multirow{2}{*}{\textbf{Discrete}}} & \multicolumn{4}{c|}{\textbf{Continuous}}                                                      & \multicolumn{2}{c|}{\multirow{2}{*}{\textbf{Discrete}}} & \multicolumn{4}{c|}{\textbf{Continuous}}                                                      \\ \cline{4-7} \cline{10-13} 
                                                & \multicolumn{2}{c|}{}                                   & \multicolumn{2}{c|}{\textbf{Blip-based}}      & \multicolumn{2}{c|}{\textbf{Vote-based}}      & \multicolumn{2}{c|}{}                                   & \multicolumn{2}{c|}{\textbf{Blip-based}}      & \multicolumn{2}{c|}{\textbf{Vote-based}}      \\ \hline
\textbf{Choice 3: Risk}                         & $R_{MSE}$                  & $R_{E[Y_d]}$               & $R_{MSE}$             & $R_{E[Y_d]}$          & $R_{MSE}$             & $R_{E[Y_d]}$          & $R_{MSE}$                  & $R_{E[Y_d]}$               & $R_{MSE}$             & $R_{E[Y_d]}$          & $R_{MSE}$             & $R_{E[Y_d]}$          \\ \hline
\textbf{Possible?}                              & \cmark      & \cmark      & \cmark & \cmark & \xmark & \cmark & \xmark      & \cmark      & \xmark & \xmark & \xmark & \cmark \\ \hline
\end{tabular}
}
\caption{Summary of the possible ODTR SuperLearner configurations across the library, metalearner, and risk dimensions. The last row (``Possible?") indicates whether the particular configuration is possible to implement. The checkmarks (\cmark) in the following table indicate that it is possible to construct that kind of ODTR SuperLearner algorithm; the x-marks (\xmark) indicate otherwise.}
\label{table0}
\end{table}
\end{center}

\section{Simulation: Comparisons and Considerations of SuperLearner ODTR Estimators}

We use simulations to: (1) illustrate the potential benefit to estimating the ODTR using a SuperLearner approach, as compared to a more traditional approach to studying treatment effect heterogeneity based on fitting an outcome regression with interaction terms on covariates and treatment, as is often standard practice; and (2) investigate the implications of practical  choices when implementing an ODTR SuperLearner in finite samples, including specification of candidate algorithms in the library, choice of metalearner, and choice of risk function.

\subsection{Data Generating Processes}

All simulations were implemented in R \citep{R}, and the code, simulated data, and results can be found at https://github.com/lmmontoya/SL.ODTR. In the future, we plan to integrate the SL.ODTR software to the Targeted Learning software ecosystem, tlverse \citep{tlverse}. We examine these comparisons using two types of data generating processes (DGPs). Each simulation consists of 1,000 iterations of either $n=1,000$ or $n = 300$, to assess the impacts of the different configurations as a function of sample size. Both DGPs generate the covariates, treatment, and outcome as follows:
\begin{align*}
W_1,W_2,W_3,W_4 &\sim Normal(\mu=0,\sigma^2=1), \\
A &\sim Bernoulli(p=0.5), \\
Y &\sim Bernoulli(p=Q_0(A,W)).\\
\end{align*}
Here, each covariate is generated from an independent Normal distribution. The probability of having a successful outcome differs for the two DGPs, which, in this case, means that the blip functions differ as well. The first DGP is an example of a complex blip function, and the second DGP is one with a blip function that is a simpler function of one variable.  The first DGP is directly from work by Luedtke and van der Laan \citep{luedtke2016super, luedtke2016optimal, van2015targeted}, and the second is modified from the first. The probability of a successful outcome for DGP 1 is:
\begin{align*}
Q_0(A,W) =& 0.5 \textrm{expit}(1-W_1^2  + 3W_2  + 5W_3^2 A - 4.45A)\\
&+0.5\textrm{expit} (-0.5- W_3  + 2W_1 W_2  + 3|W_2|A - 1.5A),
\end{align*}
then the true blip function is:
\begin{align*}
    B_0 (W)= & 0.5[\textrm{expit} (1-W_1^2  + 3W_2  + 5W_3^2  - 4.45)+\textrm{expit} (-0.5- W_3  + 2W_1 W_2  + 3|W_2|  - 1.5)\\
& - \textrm{expit} (1-W_1^2  + 3W_2 )-\textrm{expit} (-0.5- W_3  + 2W_1 W_2 )],
\end{align*}
For DGP 1, the true value of the true optimal rule $\mathbb{E}_{P_{U,X}} [Y_{d^*} ] \approx 0.5626$ and the true optimal proportion treated $\mathbb{E}_{P_{U,X} } [d^* ] \approx 55.0\%$. The true value had everyone been given treatment $\mathbb{E}_{P_{U,X}} [Y_1 ] \approx 0.4638$ and the true value had everyone not been given treatment $\mathbb{E}_{P_{U,X}} [Y_0 ] \approx 0.4643$. In the Appendix, we show results for simulations for a third DGP, following DGP 1, but with dependent covariates.

DGP 2's probability of the outcome's success is:
\begin{align*}
Q_0(A,W) &= \textrm{expit}(W_1 + 0.1A + W_1A).
\end{align*}
Thus the true blip function is:
\begin{align*}
    B_0 (W)= & \textrm{expit}(W_1 + 0.1 + W_1) - \textrm{expit}(W_1).
\end{align*}
For DGP 2, $\mathbb{E}_{P_{U,X}} [Y_{d^*} ] \approx 0.5595$ and $\mathbb{E}_{P_{U,X} } [d^* ] \approx 54.0\%$; $\mathbb{E}_{P_{U,X}} [Y_1 ] \approx 0.5152$ and $\mathbb{E}_{P_{U,X}} [Y_0 ] \approx 0.5000$.

\subsection{ODTR Estimators}

For each data generating process, we consider a number of estimators of the ODTR. First, mirroring epidemiologic practice, we model the outcome as an additive function of the treatment and covariates, and interactions with the treatment and all covariates. Such an approach translates to using the following parametric model for the outcome regression:

\[ h(E[Y|A,W]) = \beta_0 + \sum_{i=1}^k \beta_iW_i + \left(\gamma_0 + \sum_{i=1}^k \gamma_iW_i\right)A ,\]

\noindent where $h(.)$ denotes a link function, and $k$ is the number of baseline covariates in $W$. Using a linear link, the following parametric model for the blip function is implied:

\[B (W) = \gamma_0 + \sum_{i=1}^k \gamma_iW_i.\]

Next, we examine the finite sample implications of the aforementioned user-supplied choices in implementing a SuperLearner estimator of the ODTR, providing guidance for practical data analysis. First, we examine the choice of library. We consider the library that only combines candidate blip estimators (``Blip only” library; i.e., a library with candidate algorithms suited for regressing $D_n$ on $W$) versus a library that has blip estimators, direct-estimation algorithms, and static treatment rules (``Full” library). The ``Blip only" libraries consist of either:
\begin{enumerate}
    \item[(a)] Algorithms based on simple parametric models (denoted ``Parametric blip algorithms"). This consisted of univariate GLMs regressing the estimated pseudo-outcome on each covariate.
    \item[(b)] Machine learning algorithms only (denoted ``ML blip algorithms"), such as \texttt{SL.glm} (generalized linear models), \texttt{SL.mean} (the average), \texttt{SL.glm.interaction} (generalized linear models with interactions between all pairs of variables), \texttt{SL.earth} (multivariate adaptive regression splines \citep{friedman1991multivariate}), \texttt{SL.nnet} (neural networks \citep{ripley1996pattern}), \texttt{SL.svm} (support vector machines \citep{chang2011libsvm}), and \texttt{SL.rpart} (recursive partitioning and regression trees \citep{breiman2017classification}) from the SuperLearner package \citep{SLpackage}
    \item[(c)] A combination of (a) and (b) above, denoted ``Parametric + ML blip algorithms"
\end{enumerate}

The ``Full” library includes other ODTR algorithms like direct-estimation methods, static rules, and other blip-based methods. Specifically, the ``Full” library includes either the ``ML blip algorithms" or ``Parametric + ML blip algorithms" from the ``Blip only” libraries above, in addition to Q-learning \citep{kosorok2015adaptive}, OWL \citep{zhao2012estimating}, residual weighted learning (RWL) \citep{zhou2017residual}, efficient augmentation and relaxation learning (EARL) \citep{zhao2019efficient}, optimal classification algorithms \citep{zhang2012estimatingclass} (the latter 4 are from the DynTxRegime package \citep{DynTxRegime}, with function names \texttt{owl, rwl, earl,} and \texttt{optimalClass}, respectively), and static rules that treat all patients and no patients, regardless of the patient covariate profiles. For the algorithms from the DynTxRegime package, except for nuisance parameters $Q_n$ and $g_n$, we use default parameters, and the rule as a function of all covariates. Additionally, for the optimal classification algorithm, the solver method is recursive partitioning for regression trees (\texttt{rpart}). Thus, the possible ``Full" libraries are:

\begin{enumerate}
    \item[(d)] Algorithms from the ``ML blip algorithms" library, plus direct maximizers and static rules, denoted ``ML blip algorithms and $E[Y_d]$ maximizers"
    \item[(e)] All possible algorithms -- that is, algorithms from the ``Parametric + ML blip algorithms" library, plus direct maximizers and static rules, denoted ``All blip algorithms and $E[Y_d]$ maximizers"
\end{enumerate}

Second, we examine the performance of different metalearners for combining the candidate ODTR algorithms. We examine the blip-based metalearner using the ``Blip only” libraries, and the discrete and vote-based metalearners using both the ``Blip only” libraries and ``Full” libraries. 

Third, we examine the performance of ODTR estimators that use either the MSE $R_{MSE}$ or the expected outcome under the candidate rule $R_{E[Y_d]}$ as risk criteria for choosing the optimal linear combination of candidate ODTR algorithms. In particular, CV-TMLE is used for estimating $R_{E[Y_d]}$.


We fully estimate the ODTR SuperLearner by additionally estimating nuisance parameters (as opposed to using the true nuisance parameter functions) in a nested fashion \citep{coyle2017computational} as described above. Specifically, we estimate $Q_n$ and $g_n$ using the canonical SuperLearner \citep{polley2010super, SLpackage} and a correctly specified parametric regression (specifically, a logistic regression of $A$ on the intercept), respectively. We use 10-fold cross-validation throughout.

\subsection{Performance Metrics}
We measure performance by computing the percent accuracy of the algorithm; that is, the mean across simulation repetitions of the proportion of individuals for whom the treatment assigned by the estimated ODTR matches the true optimal treatment (i.e., the treatment that would have been assigned under the true ODTR). We also evaluate performance metrics of the difference between the true conditional expected counterfactual outcome under the estimated rule, averaged across the sample, compared to the true expected counterfactual outcome under the true optimal rule  $\mathbb{E}_n[Q_0(Y|A=d_n^*,W)] - \mathbb{E}_0 [Y_{d_0^*}]$ (as an approximation to the regret $\mathbb{E}_0[Q_0(Y|A=d_n^*,W)] - \mathbb{E}_0 [Y_{d_0^*}]$). We compute the mean and variance of this difference across the simulation repetitions. Instead of presenting the raw variance of the regret, we present the variance relative to that of the regret yielded by estimating the blip (and thus the optimal rule) using the ``Parametric blip algorithms" library. Additionally, we compute $2.5^{th}$ and $97.5^{th}$ quantiles of the distribution of $\mathbb{E}_n[Q_0(Y|A=d_n^*,W)]$ across the simulation repetitions. 

\subsection{Simulation Results}

Figure \ref{DGP12} displays simulation results (in addition to tables in the appendix). Below we discuss results specific to each DGP,  configuration dimension, and sample size. In general, results within each DGP (i.e., across sample sizes) follow generally similar patterns; however, any differences in performance between libraries, metalearners, or risks are more pronounced for the smaller sample size. 

\subsubsection{DGP 1 Results - ``Complex" Blip Function}

Above, we showed that DGP 1 yields a blip function that is a complex function of all of the available covariates. Here, for a larger sample size, we would expect a benefit to more aggressive approaches to estimating the ODTR, such as including more flexible machine learning-based approaches in the library of candidates, as well as use of more aggressive metalearners (either vote- or blip-based) over a discrete SuperLearner due to the better ability of these approaches to approximate the true underlying function. That said, for smaller sample sizes, this benefit might be attenuated, or even result in worse performance than simpler alternatives. For this DGP, at sample size of 1,000, indeed we find a benefit to the use of both more aggressive metalearners and larger libraries. Interestingly, this benefit is maintained for sample size 300. Specifically, libraries that included data adaptive, machine learning algorithms (as opposed to algorithms based on incorrectly specified parametric models alone) more accurately and precisely approximate the rule, even for sample size of 300. Results also show that for both sample sizes, within the discrete metalearner, the $R_{E[Y_d]}$ risk performs better than $R_{MSE}$ risk, and more saliently, the blip-based and vote-based metalearners perform better than the discrete SuperLearner. Finally, as predicted by theory, all SuperLearner approaches evaluated substantially outperform a traditional parametric regression approach at both sample sizes. Below we describe results specific to each sample size. Of note, we see results similar to those described below when there is dependence between covariates in the DGP, as shown in the Appendix.

\paragraph{$n$ = 1,000}

Libraries that contain machine learning algorithms (i.e., ``ML blip algorithms," ``Parametric + ML blip algorithms," ``ML blip algorithms and EYd maximizers, " and ``Blip algorithms and EYd maximizers") overall perform better than libraries with algorithms based on parametric models only (i.e., ``Parametric blip algorithms") and the standard GLM (i.e., ``GLM"), across all performance metrics. For example, the percent match between the true ODTR and the estimated ODTR spans from 72.0\%-77.7\% for any library with machine learning algorithms, whereas the percent match for libraries with only parametric algorithms spans from 56.4\% to 58.0\%. 

There are no stark differences within the libraries that contain machine learning algorithms across the metalearner and risk dimensions, except when using a discrete metalearner and $R_{MSE}$ risk. Specifically, the discrete metalearner that uses $R_{MSE}$ has a higher average regret and relative variance than all other algorithms that use machine learning (e.g., for the ``Parametric + ML blip algorithms" ``Blip only" library that uses a discrete metalearner, the average regret when using $R_{MSE}$ versus $R_{E[Y_d]}$ is -0.0389 versus -0.0284, respectively, and the relative variance when using $R_{MSE}$ versus $R_{E[Y_d]}$ is 2.137 versus 0.7781, respectively).

\paragraph{$n$ = 300}

As expected, given the limited data available to estimate a complex underlying function, both accuracy of treatment assignment and approximated regret (the extent to which the expected outcome under the estimated rule fell short of the best outcomes achievable) deteriorate with smaller sample sizes. That said, even in this challenging situation of a complex true pattern of treatment effect heterogeneity and limited data with which to discover it, the ODTR SuperLearner would have improved the expected outcome by approximately 4.5\% relative to the static rule that treats everyone, an approach that would have been suggested based on estimation of the ATE. 

Libraries with algorithms based on only parametric models perform worse than libraries that contain machine learning algorithms in terms of average regret and accuracy. For example, SuperLearner ODTRs that contain libraries with algorithms based on parametric blip models match 54.0\%-55.4\% of the time with the true ODTR, while the SuperLearners that contain libraries with machine learning algorithms match 60.9\%-66.1\% of the time. These results parallel those found with sample size 1,000, except the discrepancy between libraries with machine learning algorithms and parametric algorithms is not as pronounced.

Similar to the $n = 1,000$ case, among the libraries that used machine learning algorithms, using the performance of the rule as risk is better across all performance metrics than using MSE as risk for the discrete metalearner. As long as machine learning methods are included in the library, performance is similar across risk functions and choice of metalearners, with the exception of the MSE risk combined with the discrete metalearner.


\subsubsection{DGP 2 Results - ``Simple" Blip Function}

As shown above, DGP 2 has a true blip function that is a simple function of one covariate (referred to here as a ``simple" blip function). Here, the true optimal rule is described by a simple parametric model for the blip; thus, we expect this approach to perform well. However, in practice one is unlikely to be sure that the truth can be well approximated by a simple rule; it is thus of interest to evaluate what price is paid for expanding the library to include more aggressive machine learning algorithms and metalearners. In particular, one might expect that, for smaller sample size, adding machine learning-based candidates and more complex metalearners risks substantial drop-off in performance. However, specifying a library that includes algorithms based on simpler parametric models, in addition to machine learning approaches, may help mitigate this risk. In fact, for this particular DGP, we see, across metalearners and risks, only a small price in performance from adding machine learning algorithms to a library including parametric regressions. In short, having an ODTR SuperLearner library that also includes algorithms based on parametric models is better than having a library that only consists of data adaptive, machine learning algorithms. Within the libraries that did contain parametric algorithms, particularly for the discrete metalearner, using the $R_{MSE}$ risk yields slightly better performance than using the $R_{E[Y_d]}$ risk; for other metalearners there is little difference in performance in terms of risk. Performance of the metalearners varies slightly by sample size. Below we describe results specific to each sample size.

\paragraph{$n$ = 1,000}

In terms of accuracy, the libraries that only contain parametric algorithms perform the best, followed closely by libraries that contain algorithms based on parametric models and machine learning algorithms, followed by the library with only machine learning algorithms. This pattern is evident in the percent match with the true ODTR: for example, within the discrete metalearner with $R_{MSE}$ risk, the percent accuracy is 90.7\% for the library with parametric algorithms only (``Parametric blip algorithms" library), 88.8\% for the library that contain both parametric algorithms and machine learning algorithms (``Parametric + ML blip algorithms"), and 81.9\% for the library that contains machine learning algorithms only (``ML blip algorithms"). This same pattern is apparent in terms of average regret; that is, the libraries that contain algorithms based on parametric blip models or a combination of parametric and machine learning algorithms have the lowest average regret (-0.0041 to -0.0095), while the libraries that only contain machine learning algorithms have the highest average regret (-0.0100 to -0.0138). Modeling the blip with a single, parametric regression, an approach often used in standard epidemiological analysis, and, in this case, is incorrectly specified, yields an average regret of -0.0100 (higher than the libraries with a combination of parametric algorithms and/or machine learning algorithms, and at the same level as having machine learning algorithms only).

Within the libraries that contain parametric algorithms and use a discrete metalearner, using the $R_{MSE}$ risk yields better performance than the $R_{E[Y_d]}$ risk. For example, the mean regret and relative variance for the discrete metalearner that only used algorithms based on parametric models in the library is -0.0041 and 1.0267, respectively, when using $R_{MSE}$ risk, and -.0046 and 1.3174, respectively, when using $R_{E[Y_d]}$ risk. Otherwise, there were no apparent differences in performance by risk.

For libraries that contain parametric algorithms and use $R_{MSE}$, the discrete ODTR SuperLearner performs better than the blip-based ODTR SuperLearner, with regards to all performance metrics. For example, the average regret and relative variance for the library with only parametric algorithms that uses $R_{MSE}$ is -0.0041 and 1.0267, respectively, when using a discrete metalearner versus -.0059 and 1.1855, respectively, when using the blip-based metalearner. This pattern is also evident for the library that has both parametric algorithms and machine learning algorithms.

\paragraph{$n$ = 300}

As in the case where $n = 1,000$, the library with only algorithms based on parametric models performs best in terms of accuracy, followed by libraries with parametric algorithms and machine learning algorithms, and finally libraries with only machine learning algorithms; again, however, DGP 1 illustrates the risks of such a strategy. Moreover, even at this small sample size, there is only a small price to pay for adding machine learning-based learners to a library that also includes simple parametric algorithms. For example, for the discrete metalearner that uses $R_{MSE}$, the percent accuracy for the library that uses only parametric algorithms is 78.0\%, followed by a 75.5\% accuracy when there is a combination of parametric algorithms and machine learning, while the library with only machine learning algorithms had a 61.7\% match with the true ODTR. While this pattern parallels that of the $n = 1,000$ case, the dropoff in accuracy when the library uses only parametric algorithms versus when the library only uses machine learning algorithms is larger in terms of accuracy in the smaller sample size (16.3\% difference) versus the larger sample size (8.8\% difference).

Similar to the $n = 1,000$ case, among libraries that contain algorithms based on parametric models and in the discrete metalearner case, $R_{MSE}$ yields slightly better performance results than $R_{E[Y_d]}$. In contrast to the $n = 1,000$ case, for libraries that contain parametric algorithms and used $R_{MSE}$, the blip-based metalearner performs slightly better than the discrete metalearner, with regards to all performance metrics. For example, the average regret and relative variance for the library with only parametric algorithms that use $R_{MSE}$ is  -0.0188 and 1.6102, respectively, when using a blip-based metalearner versus -0.0190 and 1.8109, respectively, when using the discrete metalearner.


\section{Application of ODTR SuperLearner to ``Interventions" Study}

In the ``Interventions" study, 231 (52.2\%) participants received CBT and 210 (47.8\%) TAU. Out of the 441 participants, 271 (61.5\%) were not re-arrested within the year. The estimated probability of no re-arrest had everyone been assigned CBT is 62.2\%, and the estimated probability of no-arrest had everyone been assigned TAU is 60.7\%; there was no significant difference between these two probabilities (risk difference: 1.51\%, CI: [-8.03\%,11.06\%]). After adjusting for covariates using TMLE to improve the precision on this ATE estimate \citep{moore2009covariate}, the risk difference is, similarly, 1.53\% (CI: [-7.31\%, 10.37\%]).

\begin{figure}[h]
    \centering
    \includegraphics[scale = .37]{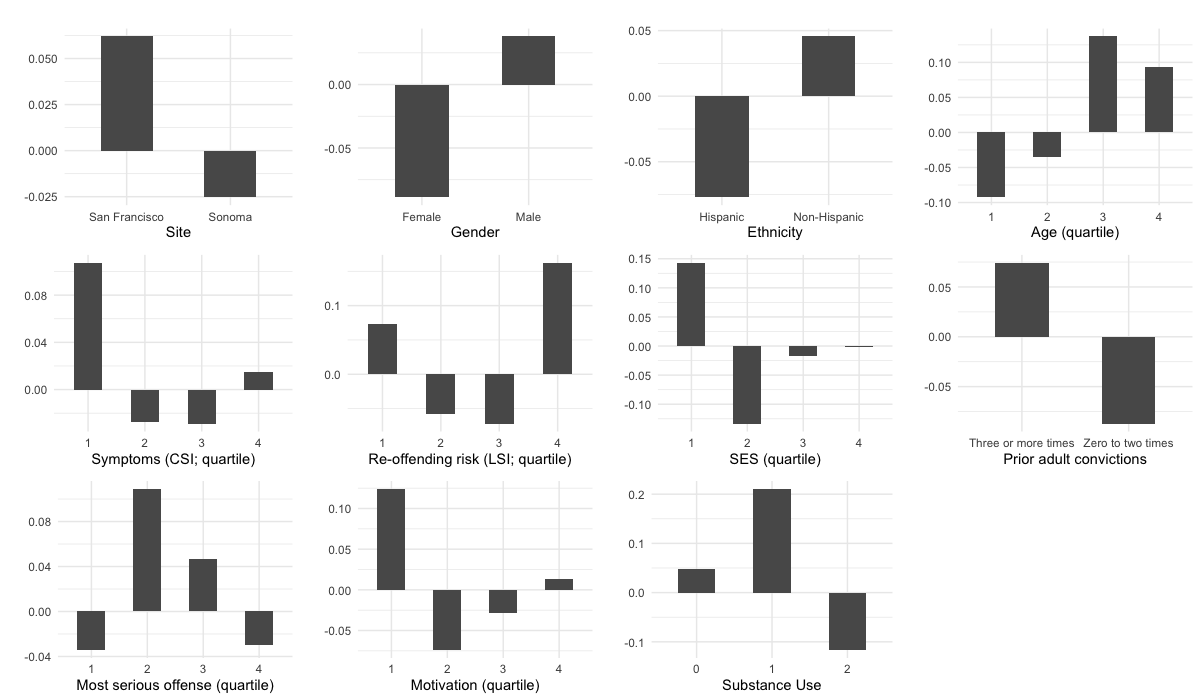}
    \caption{Subgroup plots for each of the covariates for the ``Interventions" data. The x-axis for each of the plots is the different levels of the covariates; the y-axis is the difference in proportion of people who were not re-arrested between those who received CBT versus TAU, in that covariate subgroup.}
    \label{subgroup_fig}
\end{figure}

Figure \ref{subgroup_fig} shows subgroup plots for each covariate -- that is, the unadjusted difference in probability of no re-arrest between those who received CBT versus TAU, within each covariate group level. One might begin to identify trends towards differential treatment effects; for example, participants may have benefited more from CBT at the San Francisco site, or if they had offended three or more times. Accurate interpretation of any such subgroup analyses, however, requires variance estimates and hypothesis tests with appropriate correction for multiple testing. In addition, as mentioned before, it may be the case that the best way to assign treatment is by using information on more than one variable at a time, and even interactions between those variables.

Thus, we estimated the ODTR on the ``Interventions" data to determine which justice-involved adults with mental illness should receive CBT. Specifically, we implemented the ODTR SuperLearner with a blip-only library, a continuous, blip-based metalearner, and $R_{E[Y_d]}$ risk function. We chose a blip-based library in order to generate estimates of the blip, which themselves can be informative. The library for $d^*_n$ consisted of a combination of simple parametric regressions (univariate GLMs with each covariate) and machine learning algorithms (\texttt{SL.glm}, \texttt{SL.mean}, \texttt{SL.glm.interaction}, \texttt{SL.earth}, and \texttt{SL.rpart}). As in the simulations, the outcome regression $Q_n$ was estimated using the canonical SuperLearner, $g_n$ was estimated as an intercept-only logistic regression, and we used 10-fold cross validation.

\begin{figure}[h]
    \centering
    \includegraphics[width=1\textwidth]{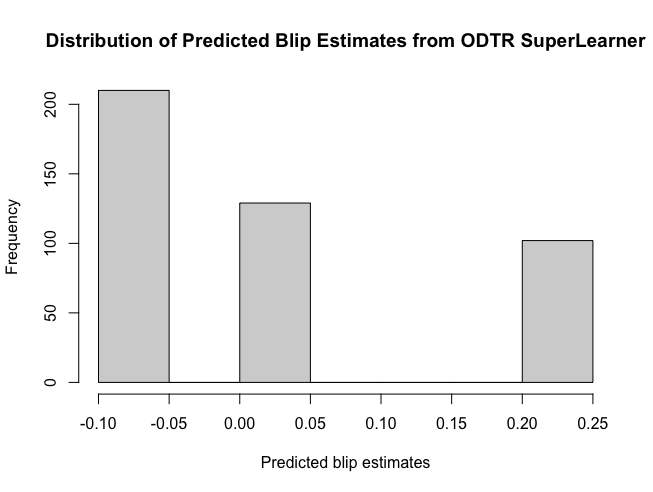}
    \caption{Distribution of predicted blip estimates from the ODTR SuperLearner. The frequencies are divided into three groups because the ODTR SuperLearner allocated all coefficient weights to a GLM using substance use, a variable with only 3 treatment levels. One can interpret the ODTR SuperLearner for this sample as follows: CBT may reduce the probability of re-arrest among justice-involved adults with low levels of substance use. Estimation and inference of the value of the ODTR SuperLearner compared to, for example, treating everyone or no one, informs us if there is, in fact, a differential effect by substance use, and thus a benefit to assigning CBT in this individualized way.}
    \label{blip_plot}
\end{figure}

Interestingly, despite implementing a continuous metalearner, the ODTR SuperLearner algorithm assigned all weight on a GLM that modeled the blip as a linear function of only substance use. As shown in Figure \ref{blip_plot}, a plot depicting the distribution of the predicted blip estimates for all participants, the algorithm can be interpreted as such: if a justice-involved person with mental illness has a low substance use score, give him/her CBT; otherwise, give him/her TAU. Under this ODTR estimate, for this sample, 52.38\% of the participants would receive CBT.

\section{Conclusions}

We described the ODTR SuperLearner and illustrated its performance for sample DGPs under different configurations of the algorithm and finite sample sizes. These results build on existing work \citep{luedtke2016super, coyle2017computational} by fully estimating the ODTR and including an expanded SuperLearner library with not only blip-based regression estimators, but also direct-estimation methods and static interventions. We highlighted the three categories of practical choices one must consider when implementing the ODTR SuperLearner: (1) the ODTR SuperLearner library of candidate algorithms, namely, whether to include parametric regression algorithms, machine learning algorithms, or both, and whether to include only estimators that output a predicted blip or include a combination of blip estimators, direct estimators, and static treatment rules; (2) the metalearner used for combining candidate algorithms, namely, whether to restrict selection to a single algorithm or combine the algorithms; and, (3) the risk function used to choose the ``best" estimator or combination of estimators of the candidate ODTR algorithms.

Simulation-based results illustrate the shortcomings of an approach to treatment effect heterogeneity based on approximating the blip as an additive function of the available covariates, or equivalently, modeling the outcome as an additive function of the treatment and covariates, and interactions between the treatment and all covariates, which is common practice in epidemiologic analyses for heterogeneous treatment effects \citep{dahabreh2016using, varadhan2013framework, yusuf1991analysis, van2015estimates}. With respect to choice of library, we recommend specifying a library with both simple parametric algorithms and more aggressive data adaptive algorithms, as well as static rules such as treat all or treat none, allowing for flexible estimation of both simple and complex underlying rules. Inclusion of a full range of algorithms from simple to aggressive was particularly important for small sample sizes. In terms of the choice of metalearner, both vote- and blip-based ensemble learners performed well; a vote-based metalearner has the advantage, however, of allowing for the integration of a larger library of candidate algorithms (including direct estimation approaches) and ease of integration of static rules. Of note, in these simulations, vote- and blip-based metalearners outperformed the discrete ODTR SuperLearner approach, even for sample size 300. However, we caution that this may not always be the case and when sample size is small, a discrete SuperLearner approach may provide benefits -- in fact, one could include a convex metalearner as a candidate algorithm. Finally, with respect to choice of risk function, both MSE and the expected outcome under the rule performed well; in practice one might prefer $R_{E[Y_d]}$ because it allows for the use of a larger library of candidate algorithms.


Additionally, as an illustration of how to apply the ODTR SuperLearner to real data, we estimated the ODTR using the ``Interventions" study to determine which types of justice-involved adults with mental illness should be assigned CBT versus TAU, to yield the highest probability of no re-arrest. Preliminary results show that the ODTR SuperLearner placed all weight on an algorithm based on a simple parametric model with only substance use; thus, the algorithm suggests that, in this sample, participants with lower levels of substance use may benefit more from CBT. We note that this is an example of a case in which the ODTR SuperLearner generated a ODTR estimate that was fully interpretable -- although we used a continuous metalearner and thus the SuperLearner could have allowed for combinations of algorithms, the SuperLearner happened to only choose one algorithm: a GLM with substance use as a regressor. To guarantee interpretability in the SuperLearner (for example, if working with practitioners who may want a treatment decision rule that could be easily written down \citep{kosorok2019precision, cohen2018treatment}), one could implement the ODTR SuperLearner with a discrete metalearner and a simple parametric library only. 


Importantly, in a companion paper, we \emph{evaluate} this ODTR -- that is, we ask the causal question: what would have been the probability of no re-arrest had participants been assigned CBT according the ODTR SuperLearner (i.e., using only substance use)? Further, is assigning CBT according to the ODTR SuperLearner significantly better than assigning CBT to everyone or no one? In this way, we can determine if it is of clinical significance to assign CBT according to this rule -- namely, if assigning CBT using only substance use scores results in a statistically significant reduction of recidivism, and if so, how much better one does with this ODTR compared to a non-individualized rule (such as giving CBT to all). Under the appropriate causal assumptions, one could use any of the methods for estimating treatment specific means to interpret this estimate as the expected outcome under the true optimal rule or the estimated optimal rule. 

Future work could extend these simulations to the multiple treatment (i.e., more than 2 treatment levels) \citep{coyle2017computational} and multiple timepoint setting (i.e., estimating a sequential ODTR from, for example, a SMART design \citep{lei2012smart, kosorok2015adaptive, almirall2014introduction} instead of an RCT design). We also plan to apply the ODTR SuperLearner on the full ``Interventions" dataset (n = 720), once data collection is finished.


This work contributes to understanding the toolbox of methods for analyzing the heterogeneity in how patients respond to treatment. By learning which patients respond best to what treatment in a flexible manner, we can improve patient outcomes -- moving us closer to the goals of precision health.

\section{Acknowledgments}

Research reported in this publication was supported by the National Institute Of Allergy And Infectious Diseases of the National Institutes of Health under Award Number R01AI074345 and F31AI140962. The content is solely the responsibility of the authors and does not necessarily represent the official views of the National Institutes of Health.

\begin{table}
\begin{tabular}{l|l|l}
\hline
  & TAU ($A=0$) & CBT ($A=1$) \\
\hline
$n$ & 211 & 230    \\
\hline
\textbf{No re-arrest} ($Y=1$) (\%) & 128 (60.7) & 143 (62.2) \\
\hline
\textbf{Site} = San Francisco (\%) & 87 (41.2) & 104 (45.2) \\
\hline
\textbf{Gender} = Female (\%) & 38 (18.0) & 37 (16.1) \\
\hline
\textbf{Ethnicity} = Hispanic (\%) & 50 (23.7) & 42 (18.3) \\
\hline
\textbf{Age} (mean (SD)) & 38.08 (11.05) & 37.01 (11.22) \\
\hline
\textbf{CSI} (mean (SD)) & 32.35 (11.13) & 33.46 (11.27)  \\
\hline
\textbf{LSI} (mean (SD)) & 5.59 (1.33) & 5.50 (1.48)  \\
\hline
\textbf{SES} (mean (SD)) & 3.81 (1.89) & 3.81 (2.12)  \\
\hline
\textbf{Prior adult convictions} (\%) &  &   \\
\hline
\hspace{3mm} Zero to two times & 74 (35.1) & 93 (40.4) \\
\hline
\hspace{3mm} Three or more times & 134 (63.5) & 129 (56.1)  \\
\hline
\hspace{3mm} Missing & 3 (1.4) & 8 (3.5)   \\
\hline
\textbf{Most serious offense} (mean (SD)) & 5.29 (2.54) & 5.09 (2.52)  \\
\hline
\textbf{Motivation} (mean (SD)) & 3.22 (1.36) & 3.27 (1.37)  \\
\hline
\textbf{Substance use} (\%) &  &   \\
\hline
\hspace{3mm} 0 & 53 (25.1) & 76 (33.0) \\
\hline
\hspace{3mm} 1 & 47 (22.3) & 55 (23.9) \\
\hline
\hspace{3mm} 2 & 109 (51.7) & 98 (42.6) \\
\hline
\hspace{3mm} Missing & 2 (0.9) & 1 (0.4)   \\
\hline
\end{tabular}
\caption{Distribution of baseline covariates in the ``Interventions" data set, stratified by randomized treatment assignment (TAU denotes Treatment as Usual, CBT denotes Cognitive Behavioral Therapy). }
\label{table1}
\end{table}

\begin{figure}[h]
    \centering
    \includegraphics[scale=.4]{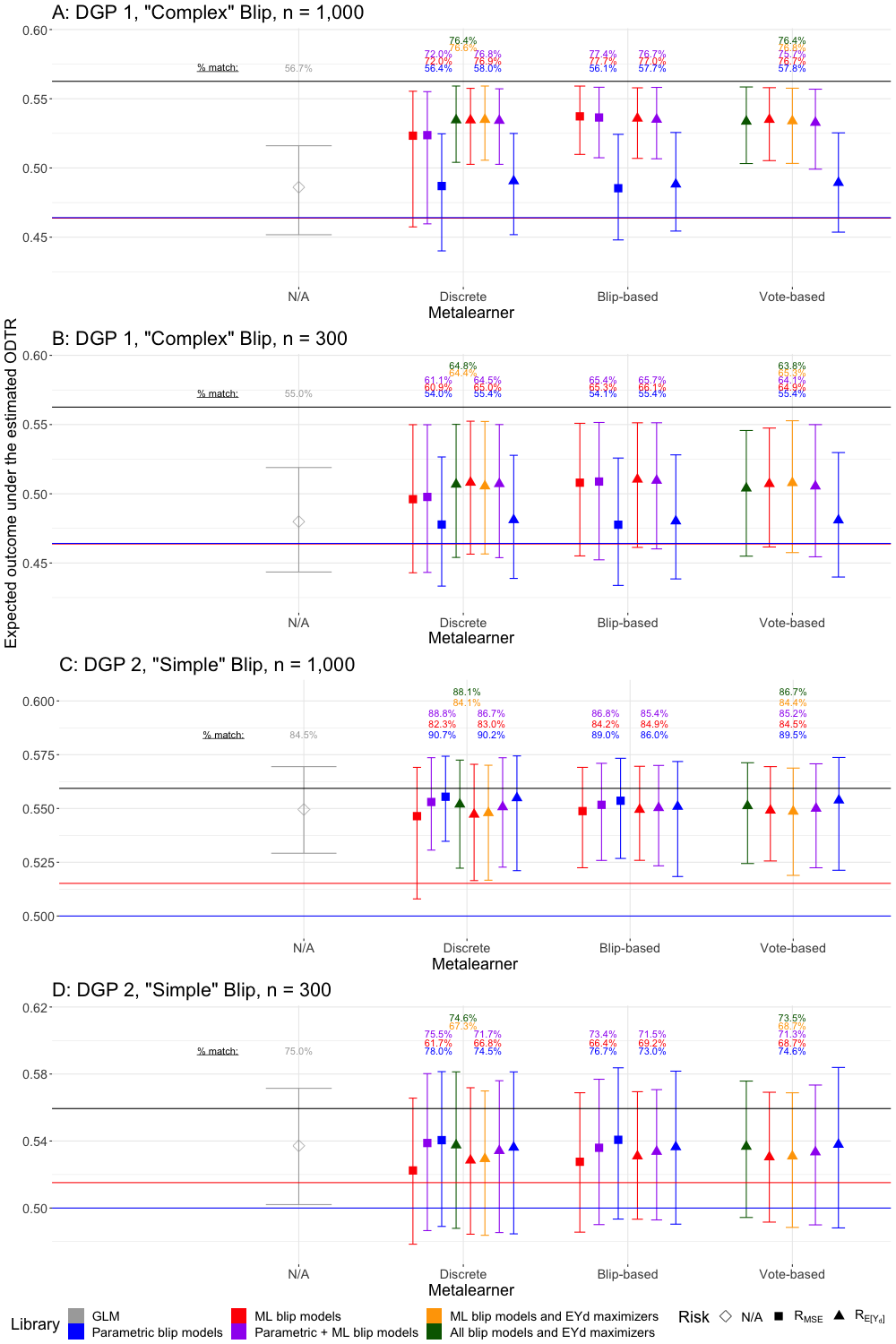}
    \caption{(Caption on the following page.)}
    \label{DGP12}
\end{figure}

\begin{figure}[t]
  \contcaption{Performance of candidate estimators of the ODTR. Plot shows mean and  $2.5^{th}$ and $97.5^{th}$ quantiles of the empirical mean of the true expected counterfactual outcome under the estimated ODTR, i.e., $\mathbb{E}_n[Q_0(Y|A=d_n^*,W)]$, an approximation of $\mathbb{E}_0[Q_0(Y|A=d_n^*(W),W)]$, for DGP 1 (top two) and DGP 2 (bottom two). The horizontal black line depicts $\mathbb{E}_{P_{U,X}}[Y_{d_0^*}]$; red line depicts $\mathbb{E}_{P_{U,X}}[Y_{1}]$; blue line depicts $\mathbb{E}_{P_{U,X}}[Y_{0}]$ (where sometimes the blue and red lines coincide and thus overlap). We compare the ODTR SuperLearner to an incorrectly specified GLM (in gray, with N/A as the metalearner and a diamond with no fill). We also compare (1) having a SuperLearner library with (a) only algorithms that estimate the blip (i.e., ``Blip only” libraries) that only have parametric algorithms (blue) or only have machine-learning blip algorithms (red) or both (purple) versus (b) an expanded or ``Full" library with blip function regressions estimated via machine learning only (orange-yellow) or machine learning and parametric algorithms (green), with methods that directly estimate the ODTR and static rules, (2) having a metalearner (depicted on the x-axis) either that chooses one algorithm (i.e., the ``discrete" SuperLearner) or combines blip predictions/treatment predictions (i.e., the ``continuous" SuperLearner), and (3) using the MSE risk function ($R_{MSE}$ as a square) versus the mean outcome under the candidate rule risk function ($R_{E[Y_d]}$ as a triangle). The percent match at the top of each plot reports the average across simulation repetitions of the percent of the sample assigned their true optimal treatment by the estimated rule.}
\end{figure}

\bibliographystyle{unsrt}
\bibliography{main}

\section{Appendix}

\subsection{Notation Table}

{\tabulinesep=0.8mm
\begin{longtabu}[]{ll}
{{\ul \textbf{Notation}}} & {\ul \textbf{Definition}} \\
\textbf{Observed Data Random Variables} &  \\
$W$ & Vector of covariates \\
$A$ & Treatment or exposure \\
$Y$ & Outcome \\
\textbf{Counterfactual outcomes} &  \\
$Y_1$ & Counterfactual outcome under treatment \\
$Y_0$ & Counterfactual outcome under control \\
$Y_d$ & Counterfactual outcome under dynamic treatment rule \\
\textbf{Models and distributions} &  \\
$\mathcal{M}^F$ & Causal model \\
$\mathcal{M}$ & Statistical model \\
$P_{U,X}$ & \begin{tabular}[c]{@{}l@{}}Probability distribution for full data $X$ (including \\ counterfactual outcomes) and unobserved \\ variables $U$. $P_{U,X} \in \mathcal{M}^F$.\end{tabular} \\
$P_0$ & \begin{tabular}[c]{@{}l@{}}True probability distribution for observed data $O$. \\ $P_{0} \in \mathcal{M}.$\end{tabular} \\
$P_n$ & \begin{tabular}[c]{@{}l@{}}Empirical probability distribution that places \\ weight $\frac{1}{n}$ on each observation\end{tabular} \\
$P_{n,v}$ & \begin{tabular}[c]{@{}l@{}}Empirical probability distribution of the validation \\ sample\end{tabular} \\
$P_{n,-v}$ & \begin{tabular}[c]{@{}l@{}}Empirical probability distribution of the training \\ sample\end{tabular} \\
\textbf{Causal parameters} &  \\
$\Psi^F_d(P_{U,X}) = \mathbb{E}_{P_{U,X}}[Y_d]$ & \begin{tabular}[c]{@{}l@{}}Expected counterfactual outcome (value) had everyone \\ received a given dynamic treatment rule $d$\end{tabular} \\
$\mathbb{E}_{P_{U,X}}[Y_1-Y_0|W]$ & Conditional average treatment effect (CATE) \\
$\Psi^F_{d_0^*}(P_{U,X}) = \mathbb{E}_{P_{U,X}} [Y_{d_0^*}]$ & \begin{tabular}[c]{@{}l@{}}Expected counterfactual outcome (value) had everyone \\ received the ODTR $d_0^*$\end{tabular} \\
$\Psi^F_{d^*_n}(P_{U,X}) = \mathbb{E}_{P_{U,X}} [Y_{d^*_n}]$ & \begin{tabular}[c]{@{}l@{}}Expected counterfactual outcome (value) had everyone \\ received the sample-specific estimate of the ODTR $d_n^*$\end{tabular} \\
$\Psi^F_{d^*_{n,v}}(P_{U,X}) = \frac{1}{V}\sum_{v=1}^V \mathbb{E}_{P_{U,X}} [Y_{d^*_{n,v}}]$ & \begin{tabular}[c]{@{}l@{}}Average of the expected validation set counterfactual \\ outcomes (values) under the training set-specific \\ estimates of the ODTR $d_{n,v}^*$\end{tabular} \\
\textbf{Statistical parameters} &  \\
$g_0(A \vert W)$ & \begin{tabular}[c]{@{}l@{}}True probability of treatment given covariates \\ (i.e., treatment mechanism; $P_0(A\vert W)$)\end{tabular} \\
$Q_0(A,W)$ & \begin{tabular}[c]{@{}l@{}}True conditional mean outcome given treatment \\ and covariates (i.e., outcome regression; $\mathbb{E}[Y\vert A,W]$)\end{tabular} \\
$d_0^*(W)$ & True optimal dynamic treatment rule (ODTR) \\
$B_0(W)$ & Blip function (identification result of the CATE) \\
$\psi_{0,d} = \Psi_{d}(P_0) =\mathbb{E}_0[Q_{0}(d(W),W)]$ & True value of a given dynamic treatment rule $d$ \\
$\psi_{0,d^*_0} = \Psi_{d^*_0}(P_0) = \mathbb{E}_0[Q_{0}(d^*_0(W),W)]$ & True value of the ODTR \\
$\psi_{0,d^*_n} \equiv \Psi_{d^*_n}(P_0) = \mathbb{E}_0[Q_{0}(d^*_n(W),W)]$ & True value of the sample-specific estimate of the ODTR \\
\begin{tabular}[c]{@{}l@{}}$\psi_{0,d^*_{n,v}} = \Psi_{d^*_{n,v}}(P_0) $\\ $=  \frac{1}{V}\sum_{v=1}^V\mathbb{E}_0[Q_{0}(d^*_{n,v}(W),W)]$\end{tabular} & \begin{tabular}[c]{@{}l@{}}Average of the true values of the training set-specific \\ estimates of the ODTR\end{tabular} \\
\textbf{Estimates} &  \\
$g_n(A \vert W)$ & Estimate of the true treatment mechanism \\
$Q_n(A,W)$ & Estimate of the outcome regression \\
$D_n(Q_n,g_n)(O)$ & \begin{tabular}[c]{@{}l@{}}Estimated pseudo-outcome. Regress this on \\ $W$ to get estimate of blip.\end{tabular} \\
$d_n^*(W)$ & ODTR estimated on the entire sample \\
$d_{n,v}^*(W)$ & Training sample-specific estimate of the ODTR \\
$d_{n,j}(W)$ & Estimate of ODTR under $j^{th}$ candidate algorithm \\
$d_{n,\alpha}(W)$ & \begin{tabular}[c]{@{}l@{}}A convex combination of candidate ODTR estimates, \\ with algorithm-specific weights given by $\alpha$\end{tabular} \\
$d_{n,B}^*(W), d_{n,d}^*$ & \begin{tabular}[c]{@{}l@{}}ODTR estimated on the entire sample  \\ using blip and direct estimation approaches, respectively\end{tabular} \\
$B_n(W)$ & Estimate of the blip \\
$B_{n,j}(W)$ & Estimate of blip under $j^{th}$ candidate algorithm \\
$B_{n,\alpha}(W)$ & \begin{tabular}[c]{@{}l@{}}A convex combination of candidate blip estimates,\\ with algorithm-specific weights given by $\alpha$\end{tabular} \\
$\hat{\psi}_d = \hat{\Psi}_d(P_n)$ & \begin{tabular}[c]{@{}l@{}}Estimate of the value of a given dynamic treatment \\ rule $d$ (estimator is either IPTW, IPTW-DR, \\ TMLE, or CV-TMLE)\end{tabular} \\
$\hat{\psi}_{d_n^*} = \hat{\Psi}_{d_n^*}(P_n)$ & \begin{tabular}[c]{@{}l@{}}Estimate of the value of sample-specific estimate \\ of the ODTR (estimator is either IPTW, \\ IPTW-DR, or TMLE)\end{tabular} \\
$\hat{\psi}_{d_{n,v}^*} = \hat{\Psi}_{d_{n,v}^*}(P_n)$ & \begin{tabular}[c]{@{}l@{}}Estimate of the value of training-set specific \\ estimate of the ODTR (estimator is CV-TMLE)\end{tabular} \\
\textbf{Risk functions} &  \\
$R_{MSE}$ & \begin{tabular}[c]{@{}l@{}}Risk for ODTR SuperLearner targeting the blip \\ function using mean-squared error\end{tabular} \\
$R_{E[Y_d]}$ & \begin{tabular}[c]{@{}l@{}}Risk for ODTR SuperLearner targeting the \\ expected rule-specific outcome\end{tabular} \\
\textbf{Other} &  \\
$d(W)$ & \begin{tabular}[c]{@{}l@{}}A given dynamic treatment rule. A function that takes \\ as input covariates $W$ (or function of $W$) and \\ outputs a treatment decision.\end{tabular} \\
$\mathcal{D}$ & Set of all dynamic treatment rules \\
$\alpha = \{\alpha_1,...,\alpha_J\}$ & \begin{tabular}[c]{@{}l@{}}Weight vector for convex combination of \\ $J$ candidate algorithms\end{tabular} \\
$\alpha_n$ & \begin{tabular}[c]{@{}l@{}}``Best" weighting of the algorithms, i.e., convex \\ combination $\alpha$ that yields the smallest cross-validated \\ empirical risk\end{tabular} \\
$\widehat{IC}(O)$ & \begin{tabular}[c]{@{}l@{}}Working influence curve of an estimator, a function \\ of the observed data $O$\end{tabular}
\end{longtabu}}

\begin{table}[]
\scalebox{0.65}{
\begin{tabular}{|l|l|l|l|l|l|l|l|l|}
\hline
\textbf{$n$}                      & \multicolumn{2}{l|}{\textbf{Library}}                                                               & \multicolumn{2}{l|}{\textbf{Metalearner}}                & \textbf{Risk} $R$   & \textbf{Avg. Regret} & \textbf{Var. Relative to GLM} & \textbf{\% Match} \\ \hline
\multirow{20}{*}{\textbf{1,000}} & \multirow{16}{*}{\textbf{Blip only}} & \textbf{GLM}                                                 & \multicolumn{2}{l|}{\textbf{N/A}}                                           & \textbf{N/A}     & -0.0765              & 1.0000 (Var = 0.0001)                       & 56.7              \\ \cline{3-9} 
                                &                                      & \multirow{5}{*}{\textbf{Parametric blip models}}             & \multicolumn{2}{l|}{\multirow{2}{*}{\textbf{Discrete}}}                     & \textbf{$MSE$}  & -0.0757              & 2.2119                        & 56.4              \\ \cline{6-9} 
                                &                                      &                                                              & \multicolumn{2}{l|}{}                                                       & \textbf{$E[Y_d]$} & -0.0721              & 1.8161                        & 58.0              \\ \cline{4-9} 
                                &                                      &                                                              & \multirow{3}{*}{\textbf{Continuous}} & \multirow{2}{*}{\textbf{Blip-based}} & \textbf{$MSE$}  & -0.0773              & 1.7008                        & 56.1              \\ \cline{6-9} 
                                &                                      &                                                              &                                      &                                      & \textbf{$E[Y_d]$} & -0.0744              & 1.6762                        & 57.7              \\ \cline{5-9} 
                                &                                      &                                                              &                                      & \textbf{Vote-based}                  & \textbf{$E[Y_d]$} & -0.0733              & 1.6635                        & 57.8              \\ \cline{3-9} 
                                &                                      & \multirow{5}{*}{\textbf{ML blip models}}                     & \multicolumn{2}{l|}{\multirow{2}{*}{\textbf{Discrete}}}                     & \textbf{$MSE$}  & -0.0393              & 2.4105                        & 72.0              \\ \cline{6-9} 
                                &                                      &                                                              & \multicolumn{2}{l|}{}                                                       & \textbf{$E[Y_d]$} & -0.0281              & 0.7664                        & 76.9              \\ \cline{4-9} 
                                &                                      &                                                              & \multirow{3}{*}{\textbf{Continuous}} & \multirow{2}{*}{\textbf{Blip-based}} & \textbf{$MSE$}  & -0.0253              & 0.5708                        & 77.7              \\ \cline{6-9} 
                                &                                      &                                                              &                                      &                                      & \textbf{$E[Y_d]$} & -0.0268              & 0.6024                        & 77.0              \\ \cline{5-9} 
                                &                                      &                                                              &                                      & \textbf{Vote-based}                  & \textbf{$E[Y_d]$} & -0.0277              & 0.7006                        & 76.7              \\ \cline{3-9} 
                                &                                      & \multirow{5}{*}{\textbf{Parametric + ML blip models}}        & \multicolumn{2}{l|}{\multirow{2}{*}{\textbf{Discrete}}}                     & \textbf{$MSE$}  & -0.0389              & 2.1327                        & 72.0              \\ \cline{6-9} 
                                &                                      &                                                              & \multicolumn{2}{l|}{}                                                       & \textbf{$E[Y_d]$} & -0.0284              & 0.7781                        & 76.8              \\ \cline{4-9} 
                                &                                      &                                                              & \multirow{3}{*}{\textbf{Continuous}} & \multirow{2}{*}{\textbf{Blip-based}} & \textbf{$MSE$}  & -0.0262              & 0.6493                        & 77.4              \\ \cline{6-9} 
                                &                                      &                                                              &                                      &                                      & \textbf{$E[Y_d]$} & -0.0276              & 0.6351                        & 76.7              \\ \cline{5-9} 
                                &                                      &                                                              &                                      & \textbf{Vote-based}                  & \textbf{$E[Y_d]$} & -0.0299              & 0.8021                        & 75.7              \\ \cline{2-9} 
                                & \multirow{4}{*}{\textbf{Full}}       & \multirow{2}{*}{\textbf{ML blip models and EYd maximizers}}  & \multicolumn{2}{l|}{\textbf{Discrete}}                                      & \textbf{$E[Y_d]$} & -0.0277              & 0.7844                        & 76.6              \\ \cline{4-9} 
                                &                                      &                                                              & \textbf{Continuous}                  & \textbf{Vote-based}                  & \textbf{$E[Y_d]$} & -0.0288              & 0.7463                        & 76.8              \\ \cline{3-9} 
                                &                                      & \multirow{2}{*}{\textbf{All blip models and EYd maximizers}} & \multicolumn{2}{l|}{\textbf{Discrete}}                                      & \textbf{$E[Y_d]$} & -0.0281              & 0.8083                        & 76.4              \\ \cline{4-9} 
                                &                                      &                                                              & \textbf{Continuous}                  & \textbf{Vote-based}                  & \textbf{$E[Y_d]$} & -0.0290              & 0.7772                        & 76.4              \\ \hline
\multirow{20}{*}{\textbf{300}}  & \multirow{16}{*}{\textbf{Blip only}} & \textbf{GLM}                                                 & \multicolumn{2}{l|}{\textbf{N/A}}                                           & \textbf{N/A}     & -0.0827              & 1.0000 (Var = 0.0003)                       & 55.0              \\ \cline{3-9} 
                                &                                      & \multirow{5}{*}{\textbf{Parametric blip models}}             & \multicolumn{2}{l|}{\multirow{2}{*}{\textbf{Discrete}}}                     & \textbf{$MSE$}  & -0.0849              & 1.5846                        & 54.0              \\ \cline{6-9} 
                                &                                      &                                                              & \multicolumn{2}{l|}{}                                                       & \textbf{$E[Y_d]$} & -0.0815              & 1.5871                        & 55.4              \\ \cline{4-9} 
                                &                                      &                                                              & \multirow{3}{*}{\textbf{Continuous}} & \multirow{2}{*}{\textbf{Blip-based}} & \textbf{$MSE$}  & -0.0850              & 1.5509                        & 54.1              \\ \cline{6-9} 
                                &                                      &                                                              &                                      &                                      & \textbf{$E[Y_d]$} & -0.0824              & 1.5346                        & 55.4              \\ \cline{5-9} 
                                &                                      &                                                              &                                      & \textbf{Vote-based}                  & \textbf{$E[Y_d]$} & -0.0817              & 1.6227                        & 55.4              \\ \cline{3-9} 
                                &                                      & \multirow{5}{*}{\textbf{ML blip models}}                     & \multicolumn{2}{l|}{\multirow{2}{*}{\textbf{Discrete}}}                     & \textbf{$MSE$}  & -0.0665              & 2.4337                        & 60.9              \\ \cline{6-9} 
                                &                                      &                                                              & \multicolumn{2}{l|}{}                                                       & \textbf{$E[Y_d]$} & -0.0545              & 1.5571                        & 65.0              \\ \cline{4-9} 
                                &                                      &                                                              & \multirow{3}{*}{\textbf{Continuous}} & \multirow{2}{*}{\textbf{Blip-based}} & \textbf{$MSE$}  & -0.0546              & 1.5906                        & 65.3              \\ \cline{6-9} 
                                &                                      &                                                              &                                      &                                      & \textbf{$E[Y_d]$} & -0.0522              & 1.3185                        & 66.1              \\ \cline{5-9} 
                                &                                      &                                                              &                                      & \textbf{Vote-based}                  & \textbf{$E[Y_d]$} & -0.0555              & 1.3437                        & 64.9              \\ \cline{3-9} 
                                &                                      & \multirow{5}{*}{\textbf{Parametric + ML blip models}}        & \multicolumn{2}{l|}{\multirow{2}{*}{\textbf{Discrete}}}                     & \textbf{$MSE$}  & -0.0649              & 2.4343                        & 61.1              \\ \cline{6-9} 
                                &                                      &                                                              & \multicolumn{2}{l|}{}                                                       & \textbf{$E[Y_d]$} & -0.0555              & 1.6545                        & 64.5              \\ \cline{4-9} 
                                &                                      &                                                              & \multirow{3}{*}{\textbf{Continuous}} & \multirow{2}{*}{\textbf{Blip-based}} & \textbf{$MSE$}  & -0.0538              & 1.6002                        & 65.4              \\ \cline{6-9} 
                                &                                      &                                                              &                                      &                                      & \textbf{$E[Y_d]$} & -0.0531              & 1.3822                        & 65.7              \\ \cline{5-9} 
                                &                                      &                                                              &                                      & \textbf{Vote-based}                  & \textbf{$E[Y_d]$} & -0.0572              & 1.5197                        & 64.1              \\ \cline{2-9} 
                                & \multirow{4}{*}{\textbf{Full}}       & \multirow{2}{*}{\textbf{ML blip models and EYd maximizers}}  & \multicolumn{2}{l|}{\textbf{Discrete}}                                      & \textbf{$E[Y_d]$} & -0.0571              & 1.6907                        & 64.4              \\ \cline{4-9} 
                                &                                      &                                                              & \textbf{Continuous}                  & \textbf{Vote-based}                  & \textbf{$E[Y_d]$} & -0.0548              & 1.5835                        & 65.3              \\ \cline{3-9} 
                                &                                      & \multirow{2}{*}{\textbf{All blip models and EYd maximizers}} & \multicolumn{2}{l|}{\textbf{Discrete}}                                      & \textbf{$E[Y_d]$} & -0.0559              & 1.6624                        & 64.8              \\ \cline{4-9} 
                                &                                      &                                                              & \textbf{Continuous}                  & \textbf{Vote-based}                  & \textbf{$E[Y_d]$} & -0.0587              & 1.5285                        & 63.8              \\ \hline
\end{tabular}
}
\caption{DGP 1 (``complex blip") results: Performance metrics (average, relative variance) of the approximate regret $E_n [Q_0(Y|A= d_n^*,W)]-E_0 [Y_{d_0^*}]$ (the difference between the average true conditional mean outcome under the estimated ODTR versus the true ODTR) for the SuperLearners generated by DGP 1. Percent agreement between the treatment assigned under the true versus estimated ODTR.}
\label{table3}
\end{table}

\begin{table}[]
\scalebox{0.65}{
\begin{tabular}{|l|l|l|l|l|l|l|l|l|}
\hline
\textbf{$n$}                      & \multicolumn{2}{l|}{\textbf{Library}}                                                               & \multicolumn{2}{l|}{\textbf{Metalearner}}                                   & \textbf{Risk} $R$   & \textbf{Avg. Regret} & \textbf{Var. Relative to GLM} & \textbf{\% Match} \\ \hline
\multirow{20}{*}{\textbf{1,000}} & \multirow{16}{*}{\textbf{Blip only}} & \textbf{GLM}                                                 & \multicolumn{2}{l|}{\textbf{N/A}}                                           & \textbf{N/A}     & -0.0100              & 1.0000 (Var = 0.0003)                       & 84.5              \\ \cline{3-9} 
                                &                                      & \multirow{5}{*}{\textbf{Parametric blip models}}             & \multicolumn{2}{l|}{\multirow{2}{*}{\textbf{Discrete}}}                     & \textbf{$MSE$}  & -0.0041              & 1.0267                        & 90.7              \\ \cline{6-9} 
                                &                                      &                                                              & \multicolumn{2}{l|}{}                                                       & \textbf{$E[Y_d]$} & -0.0046              & 1.3174                        & 90.2              \\ \cline{4-9} 
                                &                                      &                                                              & \multirow{3}{*}{\textbf{Continuous}} & \multirow{2}{*}{\textbf{Blip-based}} & \textbf{$MSE$}  & -0.0059              & 1.1855                        & 89.0              \\ \cline{6-9} 
                                &                                      &                                                              &                                      &                                      & \textbf{$E[Y_d]$} & -0.0086              & 1.5630                        & 86.0              \\ \cline{5-9} 
                                &                                      &                                                              &                                      & \textbf{Vote-based}                  & \textbf{$E[Y_d]$} & -0.0057              & 1.3902                        & 89.5              \\ \cline{3-9} 
                                &                                      & \multirow{5}{*}{\textbf{ML blip models}}                     & \multicolumn{2}{l|}{\multirow{2}{*}{\textbf{Discrete}}}                     & \textbf{$MSE$}  & -0.0138              & 1.8439                        & 81.9              \\ \cline{6-9} 
                                &                                      &                                                              & \multicolumn{2}{l|}{}                                                       & \textbf{$E[Y_d]$} & -0.0119              & 1.6144                        & 83.8              \\ \cline{4-9} 
                                &                                      &                                                              & \multirow{3}{*}{\textbf{Continuous}} & \multirow{2}{*}{\textbf{Blip-based}} & \textbf{$MSE$}  & -0.0107              & 1.3294                        & 84.2              \\ \cline{6-9} 
                                &                                      &                                                              &                                      &                                      & \textbf{$E[Y_d]$} & -0.0100              & 1.1477                        & 84.9              \\ \cline{5-9} 
                                &                                      &                                                              &                                      & \textbf{Vote-based}                  & \textbf{$E[Y_d]$} & -0.0104              & 1.1776                        & 84.5              \\ \cline{3-9} 
                                &                                      & \multirow{5}{*}{\textbf{Parametric + ML blip models}}        & \multicolumn{2}{l|}{\multirow{2}{*}{\textbf{Discrete}}}                     & \textbf{$MSE$}  & -0.0065              & 1.1875                        & 88.8              \\ \cline{6-9} 
                                &                                      &                                                              & \multicolumn{2}{l|}{}                                                       & \textbf{$E[Y_d]$} & -0.0088              & 1.4343                        & 86.7              \\ \cline{4-9} 
                                &                                      &                                                              & \multirow{3}{*}{\textbf{Continuous}} & \multirow{2}{*}{\textbf{Blip-based}} & \textbf{$MSE$}  & -0.0078              & 1.2204                        & 86.8              \\ \cline{6-9} 
                                &                                      &                                                              &                                      &                                      & \textbf{$E[Y_d]$} & -0.0092              & 1.3069                        & 85.4              \\ \cline{5-9} 
                                &                                      &                                                              &                                      & \textbf{Vote-based}                  & \textbf{$E[Y_d]$} & -0.0095              & 1.4221                        & 85.2              \\ \cline{2-9} 
                                & \multirow{4}{*}{\textbf{Full}}       & \multirow{2}{*}{\textbf{ML blip models and EYd maximizers}}  & \multicolumn{2}{l|}{\textbf{Discrete}}                                      & \textbf{$E[Y_d]$} & -0.0116              & 1.6804                        & 84.1              \\ \cline{4-9} 
                                &                                      &                                                              & \textbf{Continuous}                  & \textbf{Vote-based}                  & \textbf{$E[Y_d]$} & -0.0109              & 1.3875                        & 84.4              \\ \cline{3-9} 
                                &                                      & \multirow{2}{*}{\textbf{All blip models and EYd maximizers}} & \multicolumn{2}{l|}{\textbf{Discrete}}                                      & \textbf{$E[Y_d]$} & -0.0076              & 1.4027                        & 88.1              \\ \cline{4-9} 
                                &                                      &                                                              & \textbf{Continuous}                  & \textbf{Vote-based}                  & \textbf{$E[Y_d]$} & -0.0083              & 1.2160                        & 86.7              \\ \hline
\multirow{20}{*}{\textbf{300}}  & \multirow{16}{*}{\textbf{Blip only}} & \textbf{GLM}                                                 & \multicolumn{2}{l|}{\textbf{N/A}}                                           & \textbf{N/A}     & -0.0224              & 1.0000 (Var = 0.0004)                       & 75.0              \\ \cline{3-9} 
                                &                                      & \multirow{5}{*}{\textbf{Parametric blip models}}             & \multicolumn{2}{l|}{\multirow{2}{*}{\textbf{Discrete}}}                     & \textbf{$MSE$}  & -0.0190              & 1.8019                        & 78.0              \\ \cline{6-9} 
                                &                                      &                                                              & \multicolumn{2}{l|}{}                                                       & \textbf{$E[Y_d]$} & -0.0233              & 2.0950                        & 74.5              \\ \cline{4-9} 
                                &                                      &                                                              & \multirow{3}{*}{\textbf{Continuous}} & \multirow{2}{*}{\textbf{Blip-based}} & \textbf{$MSE$}  & -0.0188              & 1.6102                        & 76.7              \\ \cline{6-9} 
                                &                                      &                                                              &                                      &                                      & \textbf{$E[Y_d]$} & -0.0231              & 1.8938                        & 73.0              \\ \cline{5-9} 
                                &                                      &                                                              &                                      & \textbf{Vote-based}                  & \textbf{$E[Y_d]$} & -0.0216              & 2.1003                        & 74.6              \\ \cline{3-9} 
                                &                                      & \multirow{5}{*}{\textbf{ML blip models}}                     & \multicolumn{2}{l|}{\multirow{2}{*}{\textbf{Discrete}}}                     & \textbf{$MSE$}  & -0.0371              & 1.6533                        & 61.7              \\ \cline{6-9} 
                                &                                      &                                                              & \multicolumn{2}{l|}{}                                                       & \textbf{$E[Y_d]$} & -0.0310              & 1.4898                        & 66.8              \\ \cline{4-9} 
                                &                                      &                                                              & \multirow{3}{*}{\textbf{Continuous}} & \multirow{2}{*}{\textbf{Blip-based}} & \textbf{$MSE$}  & -0.0319              & 1.4242                        & 66.4              \\ \cline{6-9} 
                                &                                      &                                                              &                                      &                                      & \textbf{$E[Y_d]$} & -0.0286              & 1.2123                        & 69.2              \\ \cline{5-9} 
                                &                                      &                                                              &                                      & \textbf{Vote-based}                  & \textbf{$E[Y_d]$} & -0.0291              & 1.2973                        & 68.7              \\ \cline{3-9} 
                                &                                      & \multirow{5}{*}{\textbf{Parametric + ML blip models}}        & \multicolumn{2}{l|}{\multirow{2}{*}{\textbf{Discrete}}}                     & \textbf{$MSE$}  & -0.0207              & 1.8830                        & 75.5              \\ \cline{6-9} 
                                &                                      &                                                              & \multicolumn{2}{l|}{}                                                       & \textbf{$E[Y_d]$} & -0.0253              & 1.7460                        & 71.7              \\ \cline{4-9} 
                                &                                      &                                                              & \multirow{3}{*}{\textbf{Continuous}} & \multirow{2}{*}{\textbf{Blip-based}} & \textbf{$MSE$}  & -0.0235              & 1.5428                        & 73.4              \\ \cline{6-9} 
                                &                                      &                                                              &                                      &                                      & \textbf{$E[Y_d]$} & -0.0258              & 1.3637                        & 71.5              \\ \cline{5-9} 
                                &                                      &                                                              &                                      & \textbf{Vote-based}                  & \textbf{$E[Y_d]$} & -0.0262              & 1.4567                        & 71.3              \\ \cline{2-9} 
                                & \multirow{4}{*}{\textbf{Full}}       & \multirow{2}{*}{\textbf{ML blip models and EYd maximizers}}  & \multicolumn{2}{l|}{\textbf{Discrete}}                                      & \textbf{$E[Y_d]$} & -0.0302              & 1.5305                        & 67.3              \\ \cline{4-9} 
                                &                                      &                                                              & \textbf{Continuous}                  & \textbf{Vote-based}                  & \textbf{$E[Y_d]$} & -0.0286              & 1.4512                        & 68.7              \\ \cline{3-9} 
                                &                                      & \multirow{2}{*}{\textbf{All blip models and EYd maximizers}} & \multicolumn{2}{l|}{\textbf{Discrete}}                                      & \textbf{$E[Y_d]$} & -0.0220              & 1.7749                        & 74.6              \\ \cline{4-9} 
                                &                                      &                                                              & \textbf{Continuous}                  & \textbf{Vote-based}                  & \textbf{$E[Y_d]$} & -0.0228              & 1.5302                        & 73.5              \\ \hline
\end{tabular}
}
\caption{DGP 2 (``simple blip") results: Performance metrics (average, relative variance) of the approximate regret $E_n [Q_0(Y|A= d_n^*,W)]-E_0 [Y_{d_0^*}]$ (the difference between the average true conditional mean outcome under the estimated ODTR versus the true ODTR) for the SuperLearners generated by DGP 2. Percent agreement between the treatment assigned under the true versus estimated ODTR.}
\label{table2}
\end{table}

\pagebreak
\subsection{Simulation Extension to DGP with Dependent Covariates}

We include extra simulations illustrating a DGP scenario where covariates are dependent, as follows:

\begin{align*}
W_1 &\sim Normal(\mu=0,\sigma^2=1) \\
W_2,W_3,W_4 &\sim \mathcal{N}\left(\mathbold{\mu} = \begin{bmatrix}
0\\
0
\end{bmatrix}, \mathbold{\Sigma} = \begin{bmatrix}
1.0 & 0.3 & 0.7\\
0.3 & 1.0 & 0.8\\
0.7 & 0.8 & 1.0
\end{bmatrix}\right) \\
A &\sim Bernoulli(p=0.5) \\
Y \sim & Bernoulli(p=Q_0(A,W)),\\
\text{where } Q_0(A,W) =& 0.5\textrm{expit} (1-W_1^2  + 3W_2  + 5W_3^2 A - 4.45A)+ \\
& 0.5\textrm{expit} (-0.5- W_3  + 2W_1 W_2  + 3|W_2|A - 1.5A),
\end{align*}
then the true blip function is:
\begin{align*}
    B_0 (W)= & 0.5[\textrm{expit} (1-W_1^2  + 3W_2  + 5W_3^2  - 4.45)+\textrm{expit} (-0.5- W_3  + 2W_1 W_2  + 3|W_2|  - 1.5)\\
& - \textrm{expit} (1-W_1^2  + 3W_2 )-\textrm{expit} (-0.5- W_3  + 2W_1 W_2 )].
\end{align*}

\noindent The true expected outcome under the true ODTR $\Psi^F_{d_0^*}(P_{U,X}) \approx 0.5595$ and the true optimal proportion treated $\mathbb{E}_{P_{U,X} } [d_0^* ] \approx 54.0\%$. The mean outcome had everyone and no one been treated are, respectively, $\mathbb{E}_{P_{U,X}} [Y_1 ] \approx 0.5152$ and $\mathbb{E}_{P_{U,X}} [Y_0 ] \approx 0.5000$.

\begin{figure}[h]
    \centering
    \includegraphics[scale=.4]{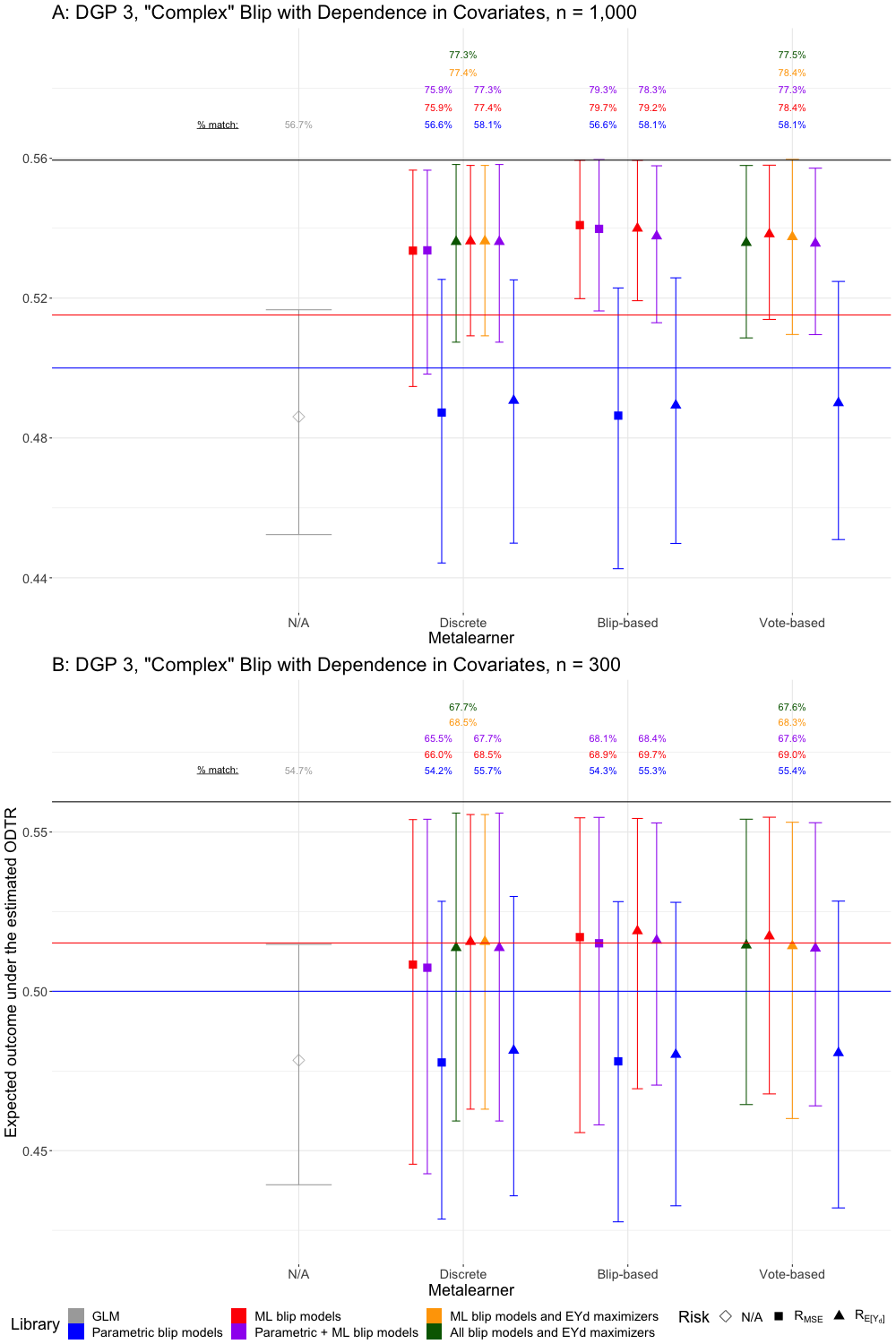}
    \caption{(Caption on the following page.)}
    \label{DGP122}
\end{figure}

\begin{figure}[t]
  \contcaption{Performance of candidate estimators of the ODTR. Plot shows mean and  $2.5^{th}$ and $97.5^{th}$ quantiles of the empirical mean of the true expected counterfactual outcome under the estimated ODTR, i.e., $\mathbb{E}_n[Q_0(Y|A=d_n^*,W)]$, an approximation of $\mathbb{E}_0[Q_0(Y|A=d_n^*(W),W)]$, for DGP 1 (top two) and DGP 2 (bottom two). The horizontal black line depicts $\mathbb{E}_{P_{U,X}}[Y_{d_0^*}]$; red line depicts $\mathbb{E}_{P_{U,X}}[Y_{1}]$; blue line depicts $\mathbb{E}_{P_{U,X}}[Y_{0}]$ (where sometimes the blue and red lines coincide and thus overlap). We compare the ODTR SuperLearner to an incorrectly specified GLM (in gray, with N/A as the metalearner and a diamond with no fill). We also compare (1) having a SuperLearner library with (a) only algorithms that estimate the blip (i.e., ``Blip only” libraries) that only have parametric algorithms (blue) or only have machine-learning blip algorithms (red) or both (purple) versus (b) an expanded or ``Full" library with blip function regressions estimated via machine learning only (orange-yellow) or machine learning and parametric algorithms (green), with methods that directly estimate the ODTR and static rules, (2) having a metalearner (depicted on the x-axis) either that chooses one algorithm (i.e., the ``discrete" SuperLearner) or combines blip predictions/treatment predictions (i.e., the ``continuous" SuperLearner), and (3) using the MSE risk function ($R_{MSE}$ as a square) versus the mean outcome under the candidate rule risk function ($R_{E[Y_d]}$ as a triangle). The percent match at the top of each plot reports the average across simulation repetitions of the percent of the sample assigned their true optimal treatment by the estimated rule.}
\end{figure}

\end{document}